\documentstyle[prd,eqsecnum,aps]{revtex}

\begin{document}

\draft
\title{Particle production in the oscillating inflation model}
\author{Shinji Tsujikawa \thanks{electronic
address:shinji@gravity.phys.waseda.ac.jp}}
\address{ Department of Physics, Waseda University,
Shinjuku, Tokyo 169-8555, Japan\\[.3em]}
\date{\today}
\maketitle
\begin{abstract}
We investigate the particle production of 
a scalar field $\chi$ coupled to an inflaton 
field $\phi$ ($g^2\phi^2\chi^2/2$) in the {\it oscillating inflation} 
model, which was recently proposed by Damour and Mukhanov.
Although the fluctuation of the $\phi$ field can be effectively 
enhanced during a stage of the oscillating inflation,
the maximum fluctuation is suppressed as the critical
value $\phi_c$ which indicates the scale of the core part 
of the inflaton potential decreases, in taking into account 
the back reaction effect of created particles. 
As for the $\chi$ particle production,
we find that larger values of the coupling constant $g$ are
required to lead to an efficient parametric resonance
with the decrease of $\phi_c$,
because an effective mass of inflaton 
around the minimum of its potential becomes larger.
However, it is possible to generate the superheavy $\chi$ particle 
whose mass is greater than $10^{14}$ GeV,
which would result in an important consequence 
for the GUT baryogenesis.
\end{abstract}


\baselineskip = 18pt

%
\section{Introduction}                            %
Inflationary cosmology is one of the most reliable concepts 
to describe the early stage of the universe\cite{GS}.
This paradigm not only gives the solution to a number of shortcomings
of the standard big bang cosmology, but also provides the density 
perturbation that may be responsible for the structure
formation\cite{Kolb}.
During the inflationary stage, a scalar field $\phi$ known as 
{\it inflaton} is slowly rolling down toward a minimum 
of its potential. Inflation ends when the kinetic energy of the inflaton
becomes comparable to the potential energy. 
After that, the inflaton field begins to oscillate
around the minimum of its potential and
produces elementary particles.
This process by which the energy of the inflaton field is  
transferred to other particles is called reheating.
The original scenario of reheating was considered in Ref.
~\cite{DA} based on the perturbation theory, adding the 
phenomenological decay term to the equation of the inflaton field.
However, since the production of a grand unified theory (GUT)
scale boson is kinematically forbidden in this model, the GUT 
scale baryogenesis does not work well in this scenario.

It was recently recognized that the reheating process begins
by an explosive particle production called {\it preheating}
\cite{TB,KLS1}.
During this stage, the fluctuation of produced particles 
grows quasi-exponentially by parametric resonance.
The efficiency of resonance depends on the model of the inflation.
The chaotic inflation is one of the most efficient model for the
development of the fluctuation.
In the case of the massive inflaton potential
$V(\phi)=m^2\phi^2/2$,
another scalar field $\chi$ coupled to inflaton with a coupling 
$g^2\phi^2\chi^2/2$ can be enhanced in a certain range 
of the coupling constant\cite{KT1,PR,KLS2}. 
With the coupling of $g~\mbox{\raisebox{-1.ex}{$\stackrel
     {\textstyle>}{\textstyle\sim}$}}~10^{-4}$,
resonance turns on from a broad resonance regime and 
the fluctuation of the $\chi$ particle increases overcoming the 
diluting effect by the cosmic expansion.
In the massless inflaton potential $\lambda\phi^4/4$, 
the growth of the inflaton fluctuation occurs even if we do not
introduce another field $\chi$ coupled to inflaton
\cite{Boy,Kaiser,KT2,Son,Baa}.
In this case, however, since the resonance band is restricted 
to be narrow, the transfer of energy from
the homogeneous inflaton
to the fluctuation does not occur sufficiently.
In the two-field model of $V(\phi,\chi)= \lambda\phi^4/4
+g^2\phi^2\chi^2/2$, the $\chi$ particle production is typically
more efficient than the production of the
$\phi$ particle\cite{Greene}.
As for other models of inflation, several authors considered preheating 
such as the hybrid inflation\cite{JL}, and the higher curvature 
inflation\cite{TMT}.

Recently it has been recognized by Damour and Mukhanov\cite{DM}
that inflation occurs even in the oscillating stage for the non-convex
type potential $V(\phi)$ where $d^2V/d\phi^2$ is negative
in the regime not too far from the core of the potential. 
In spite of the rapid oscillation of the inflaton field, 
inflation takes place during 
which the field moves in flat regions outside the core part.
Liddle and Mazumdar\cite{LM} called it {\it oscillating inflation} and 
numerically calculated the revised number of $e$-foldings.
The amount of inflation becomes larger with the decrease of
the critical value $\phi_c$ which determines the scale of
the convex core of the potential. 
However, it is difficult to induce a sufficient inflation
only by the oscillating inflation even when $\phi_c$ is
lowered to the electro-weak scale 
$\phi_c=10^{-17}M_{\rm pl}$\cite{LM}.
Since the ordinary inflation takes place in the slow-roll 
regime before the oscillating inflation, the total amount 
of $e$-folding is mostly due to this preceding inflation
and the contribution of the oscillating inflation
is added to some extent.
The energy scale of the potential can be determined in the usual manner
by fitting the density perturbation observed by the Cosmic Background
Explorer (COBE) satellite.

As was pointed out by Damour and Mukhanov, 
since the square of the effective
mass of the inflaton field $m_{\phi}^2 \equiv
d^2V/d\phi^2$ is mostly negative during the
oscillating inflation, the fluctuation of inflaton 
can be strongly amplified.
Taruya\cite{Taruya} considered the evolution of the fluctuation
including the metric perturbation in the single field model 
during the oscillating inflation.
Although the super-horizon modes ($k\to 0$) 
are not relevantly amplified,
the growth of the modes inside the Hubble horizon 
occurs significantly.
In the case of the single field, since the curvature perturbation 
on the comoving slice remains constant in the long 
wavelength limit, we can not expect the significant growth 
of the long-wave perturbation.
It was also suggested by Damour and Mukhanov
that another field $\chi$ coupled to inflaton would be 
resonantly amplified during the oscillating inflation
and the superheavy GUT scale bosons
are expected to be generated.
The structure of resonance for the $\chi$ particle is different from  
that of the $\phi$ particle, and we should make clear 
the parameter range of the coupling constant $g$ where 
the $\chi$ particle production occurs sufficiently.
Since the effective mass of the inflaton around the minimum of the
potential depends on the critical value $\phi_c$,
it is also important to investigate how the difference
of the shape of the potential would affect the $\chi$ particle production.
If the $\chi$ particle is efficiently created in the oscillating inflation model,
this would affect a nonthermal 
phase transition\cite{nonthermal}, and 
topological defect formation\cite{defect}.
Also, the possibility of the production of the superheavy particle
would make the baryogenesis at the GUT
scale possible\cite{baryogenesis}.
In this paper, we consider the production of the $\chi$ particle
as well as the $\phi$ particle during the oscillating inflation 
and the subsequent oscillating stage. 
We make use of the Hartree approximation in order to include 
the back reaction effect of created particles which is absent in Ref.
\cite{Taruya}.
Even in the case where the resonance band is very broad, the back 
reaction effect plays a crucial role to terminate the growth of the 
fluctuation. We will discuss how the maximum fluctuations of 
$\phi$ and $\chi$ particles depend on 
the critical value of $\phi_c$ and the coupling constant $g$.

This paper is organized as follows.
In the next section, we introduce basic equations based on the 
Hartree approximation in the oscillating inflation model.
In Sec.~III, the particle production in this model is investigated 
by making use of the numerical results. 
We study how the fluctuations of  the $\phi$ and $\chi$ fields
grow during the oscillating inflation and the subsequent 
oscillating stage.
We give our discussions and conclusions in the final section.

\section{The basic equations}   
 
We investigate a model where an inflaton field $\phi$ is coupled to
a scalar field $\chi$\cite{comment},
\begin{eqnarray}
{\cal L} = \sqrt{-g} \left[ \frac{1}{2\kappa^2}R
   -\frac12 (\nabla \phi)^2
   -V(\phi)
   -\frac12 (\nabla \chi)^2
   -\frac12 m_{\chi}^2 \chi^2
   -\frac12 g^2\phi^2\chi^2
    \right],
\label{B1}
\end{eqnarray}
where $\kappa^{2}/8\pi \equiv G =M_{\rm pl}^{-2} $ is Newton's
gravitational constant, $R$ is a scalar curvature, $m_{\chi}$ is 
a mass of the $\chi$ field, and $g$ is a coupling constant.
We consider an inflaton potential $V(\phi)$ proposed by
Damour and Mukhanov\cite{DM}, 
which is described by
\begin{eqnarray}
V(\phi)=\frac{M^4}{q} \left[ \left( \frac{\phi^2}{\phi_c^2}
+1 \right)^{q/2} -1 \right],
\label{B2}
\end{eqnarray}
where $M$ is a mass which is constrained by
the primordial density perturbation observed by the 
COBE satellite,
$q$ is a dimensionless parameter greater than zero,
and $\phi_c$ is a critical value of the inflaton field 
which determines the scale of 
the core part of the potential (See Fig.~1).
Note that the potential $(\ref{B2})$ is a toy model which is not directly 
related with a real theory of physics, although supergravity and
superstring models may give rise to some non-convex
type potentials\cite{DM}. In the case of $\phi^2 \gg \phi_c^2$, the potential is 
approximately written by $V(\phi) \approx Aq^{-1}
(\phi/\phi_c)^q$, and $d^2 V/d\phi^2$ becomes negative for $q<1$.
Inflation takes place in the usual manner while
inflaton slowly moves in this flat region.
This conventional slow-roll inflation is followed by the 
{\it oscillating inflation}, which means that inflation continues
to occur during the oscillating stage of inflaton while it evolves
in the region of $\phi^2~\mbox{\raisebox{-1.ex}{$\stackrel
     {\textstyle>}{\textstyle\sim}$}}~\phi_c^2$.
Since the amplitude of the $\phi$ field gradually decreases 
due to the adiabatic expansion of the universe,
the $\phi$ field is finally trapped in the core region
of the potential (i.e. $\phi^2~\mbox{\raisebox{-1.ex}{$\stackrel
     {\textstyle<}{\textstyle \sim}$}}~\phi_c^2$) and the oscillating
inflation ceases. After that, the universe enters the ordinary
reheating stage, in which the potential $(\ref{B2})$ is described by 
the massive inflaton potential. 

Hereafter, we study the dynamics of the system
in the flat Friedmann-Robertson-Walker metric
\begin{eqnarray}
ds^2 = -dt^2 + a^2(t) d {\bf x}^2,
\label{B7}
\end{eqnarray}
where $a(t)$ is the scale factor, and $t$ is the 
cosmic time coordinate. 

Let us consider the equations of motion based on the 
Hartree factorization in the oscillating inflation model. 
We decompose the inflaton field into the homogeneous 
and fluctuational parts as
\begin{eqnarray}
\phi(t,{\bf x})=\phi_0(t) +\delta \phi(t,{\bf x}),
\label{B8}
\end{eqnarray}
where the fluctuational part satisfies the tadpole condition
\begin{eqnarray}
\langle \delta \phi(t,{\bf x}) \rangle=0.
\label{B9}
\end{eqnarray}
In order to study the quantum particle creation,
we expand $\delta \phi$ and $\chi$ fields as
\begin{eqnarray}
\delta \phi=\frac{1}{(2\pi)^{3/2}} \int 
\left(a_k 
\delta \phi_k(t)
 e^{-i {\bf k} \cdot {\bf x}}+a_k^{\dagger} 
\delta \phi_k^{*}(t)
 e^{i {\bf k} \cdot {\bf x}} 
\right) d^3{\bf k},
\label{B10}
\end{eqnarray}
\begin{eqnarray}
\chi=\frac{1}{(2\pi)^{3/2}} \int 
\left(a_k 
\chi_k(t)
 e^{-i {\bf k} \cdot {\bf x}}+a_k^{\dagger} 
\chi_k^{*}(t)
 e^{i {\bf k} \cdot {\bf x}} 
\right) d^3{\bf k},
\label{B11}
\end{eqnarray}
where $a_k$ and $a_k^{\dagger}$ are the annihilation and creation 
operators respectively.

Then, the equations of motion for inflaton are expressed by
imposing the Hartree factorization\cite{CH} as
\begin{eqnarray}
\ddot{\phi}_0 +3H \dot{\phi}_0 +
\sum_{n=0}^{\infty} \frac{1}{2^n n!}
\langle \delta\phi^2 \rangle^n V^{(2n+1)} (\phi_0) 
+g^2 \langle\chi^2\rangle \phi_0=0,
\label{B30}
\end{eqnarray}
\begin{eqnarray}
\delta \ddot{\phi}_k +3H \delta \dot{\phi}_k+
\left[ \frac{k^2}{a^2} +\sum_{n=0}^{\infty} \frac{1}{2^n n!}
\langle \delta\phi^2 \rangle^n V^{(2n+2)} (\phi_0) 
+g^2 \langle\chi^2\rangle \right]
\delta \phi_k=0,
\label{B12}
\end{eqnarray}
where $H \equiv \dot{a}/a$, 
$V^n(\phi_0) \equiv  \delta^n V(\phi_0)/\delta \phi_0^n$,
and  $\langle \delta \phi^2 \rangle$, $\langle \chi^2 \rangle$
are defined by
\begin{eqnarray}
\langle \delta \phi^2 \rangle = \frac1{2\pi^2} \int k^2
|\delta \phi_k|^2 dk,
\label{B40}
\end{eqnarray}
\begin{eqnarray}
\langle \chi^2 \rangle = \frac1{2\pi^2} \int k^2
|\chi_k|^2 dk.
\label{B41}
\end{eqnarray}
Note that the $V^{(2)}(\phi_0)$ term in Eq.~$(\ref{B12})$ 
is the leading term for the development of the fluctuation
$\langle \delta \phi^2 \rangle$. 
In the present model, this term is negative in most 
stages of the oscillating inflation, and the growth of the fluctuation 
can be expected during this stage.
As $\phi$ particles are produced, however, 
the back reaction effect by the growth of 
$\langle \delta \phi^2 \rangle$ plays 
an important role.
The development of the fluctuation is suppressed by the 
$\phi$ particle production itself.

With regard to the $\chi$ field, this satisfies the following equation
\begin{eqnarray}
\ddot{\chi}_k +3H\dot{\chi}_k+\left[\frac{k^2}{a^2}
+m_{\chi}^2+g^2(\phi_0^2+\langle\delta\phi^2\rangle)
\right] \chi_k=0.
\label{B14}
\end{eqnarray}
The $g^2 \phi_0^2$ term leads to the parametric amplification 
of $\chi$ particles when the $\phi_0$ field oscillates around the
minimum of its potential.
The strength of resonance depends on the coupling
constant $g$ and the amplitude of the $\phi_0$ field.
Moreover, the effective mass $m_{\phi}$ of the $\phi_0$ 
field is also important as we will show later.
Since $m_{\phi}$ is closely related to the shape around 
the core region, the development of the fluctuation
$\langle\chi^2\rangle$ depends on the critical value $\phi_c$.
In the case where the growth rate of the $\chi$ fluctuation is large,
back reaction effects of $\chi$ particles as well as $\phi$
particles finally shut off the parametric resonance.

Next, the evolution of the scale factor is described by
\begin{eqnarray}
\left(\frac{\dot{a}}{a}\right)^2  & = &
   \frac{\kappa^2}{3} 
     \biggl[ \frac12 \dot{\phi_0}^2
   +\frac12 \langle \delta \dot{\phi}^2 \rangle
   +\frac{1}{2a^2} \langle (\nabla \phi )^2 \rangle 
   +\sum_{n=0}^{\infty} \frac{1}{2^n n!}
  \langle \delta\phi^2 \rangle^n V^{2n}(\phi_0)
   \nonumber \\
   & + &   
   \frac12 \langle \dot{\chi}^2 \rangle
   +\frac{1}{2a^2} \langle (\nabla \chi)^2 \rangle
  +\frac12 \left\{m_{\chi}^2+g^2(\phi_0^2+
   \langle \delta\phi^2 \rangle) \right\} \langle\chi^2\rangle
\biggr], 
\label{B100}
\end{eqnarray}
where 
\begin{eqnarray}
\langle \delta \dot{\phi}^2 \rangle=
\frac1{2\pi^2} \int k^2
|\delta \dot{\phi}_k|^2 dk,
\label{B60}
\end{eqnarray}
\begin{eqnarray}
\langle (\nabla \phi )^2 \rangle =
\frac{1}{2\pi^2} \int k^4
|\delta \phi_k|^2 dk,
\label{B61}
\end{eqnarray}
\begin{eqnarray}
\langle (\nabla \chi)^2 \rangle=
\frac1{2\pi^2} \int k^4 |\chi_k|^2 dk.
\label{B62}
\end{eqnarray}

Before analyzing the above equation of motion, 
we first discuss the evolution of the scale factor 
and the homogeneous inflaton field neglecting
the fluctuational terms. Then, Eqs.~$(\ref{B30})$ and  
$(\ref{B100})$ are approximately written as
\begin{eqnarray}
\ddot{\phi}_0 +3H \dot{\phi}_0 +
V,_{\phi_0}(\phi_0) \approx 0,
\label{B16}
\end{eqnarray}
\begin{eqnarray}
\left(\frac{\dot{a}}{a}\right)^2 \approx
   \frac{\kappa^2}{3} 
     \left[ \frac12 \dot{\phi_0}^2
    +V(\phi_0) \right]. 
\label{B15}
\end{eqnarray}
We can understand the mean behavior of the scale factor and
the inflaton field by ignoring the core region of the potential.
For the case of $\phi_0^2 \gg \phi_c^2$,
making use of the time averaged relation
$\langle\dot{\phi}^2 \rangle_T=
q \langle V(\phi) \rangle_T$ in  
Eqs.~$(\ref{B16})$ and $(\ref{B15})$,
we find the following approximate 
relation\cite{DM}
\begin{eqnarray}
a \propto t^{(q+2)/3q},
\label{B3}
\end{eqnarray}
\begin{eqnarray}
\tilde{\phi}_0 \propto t^{-2/q},
\label{B17}
\end{eqnarray}
where $\tilde{\phi}_0$ is the amplitude of the $\phi_0$ field.
Note that the universe expands acceleratedly for $0<q<1$.
Hereafter, we consider the case of $0<q<1$.
Since inflation takes place for $\phi_0^2~\mbox{\raisebox{-1.ex}{$\stackrel
     {\textstyle>}{\textstyle\sim}$}}~\phi_c^2$,
the amount of inflation becomes larger 
with the decrease of $\phi_c$.
Damour and Mukhanov estimated the number of $e$-foldings as
\begin{eqnarray}
N \equiv {\rm ln}\frac{a_f}{a_s} \approx
\frac{2+q}{6} {\rm ln}\frac{\phi_0(t_s)}{\phi_c},
\label{B4}
\end{eqnarray}
where the subscripts $s$ and $f$ denote the end of the slow-roll
and  the end of the oscillating inflation respectively.
The slow-roll inflation ends when the slow-roll parameter 
$\epsilon \equiv \left(V,_{\phi_0}/V \right)^2/2\kappa^2$
becomes of order unity. For the case of 
$\phi_0^2 \gg \phi_c^2$, since $\epsilon$ can be written as
\begin{eqnarray}
\epsilon  \approx \frac{q^2M_{\rm pl}^2}
{16\pi \phi_0^2},
\label{B5}
\end{eqnarray}
$\phi_0(t_s)$ is estimated by setting $\epsilon=1$ as
\begin{eqnarray}
\phi_0(t_s) \approx \frac{q}{\sqrt{16\pi}} M_{\rm pl}.
\label{B6}
\end{eqnarray}
We find from Eqs.~$(\ref{B4})$ and $(\ref{B6})$ that 
the number of $e$-foldings during the oscillating inflation
is small compared with the total needed amount of inflation 
$N_{t}~\mbox{\raisebox{-1.ex}{$\stackrel
     {\textstyle>}{\textstyle\sim}$}}~60$.
For example, when $\phi_c=10^{-6} M_{\rm pl}$ 
and $q=0.1$, $N=3.3$. 
Even if $\phi_c $ is the electro-weak scale
$\phi_c=10^{-17} M_{\rm pl}$, $N=12.2$ for $q=0.1$.
As was confirmed by numerical calculations 
in Ref.~\cite{LM}, the number of $e$-foldings becomes smaller
as $q$ approaches zero or unity for the fixed
value of $\phi_c$. Hence even if we change the values of $q$
in the range of $0<q<1$, 
the contribution to the amount of inflation due to  the 
oscillating inflation is still small. 
This means that the preceding inflation by the 
ordinary slow-roll is expected to contribute to most 
of the number of $e$-foldings needed 
to solve cosmological puzzles.

Let us estimate the energy scale of the potential $(\ref{B2})$
which is constrained by the density perturbation observed by
the COBE satellite.
First, we consider the value of $\phi_0$ ($=\phi_0(t_i)$)
at the epoch of horizon exit when physical scales crossed outside
the Hubble radius 50 $e$-foldings before the start of the oscillating
inflation. Calculating the number of $e$-foldings 
\begin{eqnarray}
N = -\kappa^2 \int_{\phi_0(t_i)}^{\phi_0(t_s)}
\frac{V}{V, _{\phi_0}} d\phi_0,
\label{B18}
\end{eqnarray}
in the present model with the condition 
of $\phi_0^2 \gg \phi_c^2$, we obtain
\begin{eqnarray}
N \approx \frac{4\pi}{qM_{\rm pl}^2}
\left[\phi_0^2(t_i)-\phi_0^2(t_s) \right].
\label{B19}
\end{eqnarray}
Then the value of $\phi_0 (t_i)$ is approximately estimated as
\begin{eqnarray}
\phi_0 (t_i) \approx \sqrt{\frac{qN}{4\pi}} M_{\rm pl},
\label{B20}
\end{eqnarray}
where we neglected the contribution of the $\phi_0(t_s)$ term.
The square of the amplitude of the density perturbation can be
calculated for $\phi_0^2 \gg \phi_c^2$ as\cite{Lyth}
\begin{eqnarray}
\delta_H^2 = \frac{32}{75} 
\frac{V(\phi_0(t_i))}{M_{\rm pl}^4}
\epsilon^{-1} (\phi_0(t_i))
\approx 
\frac{512\pi}{75} \frac{M^4}{q^3 M_{\rm pl}^6}
\frac{\phi_0^{q+2}(t_i)}{\phi_c^q}.
\label{B21}
\end{eqnarray}
Then the mass $M$ is constrained to be
\begin{eqnarray}
M \approx \left[ \frac{75}{128}\frac{q^2}{N}
\left(\frac{\phi_c}{M_{\rm pl}} \sqrt{\frac{4\pi}{qN}}
\right)^q \delta_H^2 \right]^{1/4} M_{\rm pl}.
\label{B22}
\end{eqnarray}
The COBE data requires $\delta_H \approx
2 \times10^{-5}$. Setting $N=50$, we can estimate
the mass $M$ as two functions of $q$ and 
$\phi_c$. The mass $M$ is weakly dependent on the 
value of $\phi_c$ for the fixed value of $q$. 
For example, for $\phi_c=10^{-4}M_{\rm pl}$
and $q=0.1$, $M=3.74 \times 10^{-4}M_{\rm pl}$; and
for $\phi_c=10^{-6}M_{\rm pl}$
and $q=0.1$, $M=3.33 \times 10^{-4}M_{\rm pl}$.
In the next section, we use the value of $M$
which is obtained by Eq.~$(\ref{B22})$,
and consider the growth of the fluctuation during and
after the oscillating inflation.

\section{Particle production in the oscillating inflation model}   

In this section, we investigate the particle production
of $\phi$ and $\chi$ fields with the potential
$(\ref{B2})$. 
Defining new scalar fields $\varphi_0 \equiv a^{3/2} \phi_0$,
$\delta \varphi_k \equiv a^{3/2} \delta \phi_k$, and 
$X_k \equiv a^{3/2} \chi_k$, Eqs.~$(\ref{B30})$,
$(\ref{B12})$, and $(\ref{B14})$ can be rewritten as 
\begin{eqnarray}
\ddot{\varphi}_0+\omega_{\varphi_0}^2
\varphi_0=0,
\label{P1}
\end{eqnarray}
\begin{eqnarray}
\delta \ddot{\varphi}_k +\omega_{\delta\varphi_k}^2
\delta \varphi_k=0,
\label{P2}
\end{eqnarray}
\begin{eqnarray}
\ddot{X}_k +\omega_{X_k}^2 X_k=0,
\label{P3}
\end{eqnarray}
with
\begin{eqnarray}
\omega_{\varphi_0}^2 & \equiv & \frac{M^4}{\phi_c^2}
\left(\frac{\phi_0^2}{\phi_c^2}+1 \right)^
{\frac12 q-1}+\frac12 \frac{M^4}{\phi_c^4}
\langle\delta\phi^2\rangle
 \left(\frac{\phi_0^2}{\phi_c^2}+1 \right)^
{\frac12 q-3} (q-2) \left\{ (q-1) \frac{\phi_0^2}{\phi_c^2}
+3 \right\}  \nonumber \\
& + & 
g^2 \langle\chi^2\rangle
-\frac34\left(\frac{2\ddot{a}}{a}
+\frac{\dot{a}^2}{a^2}\right),
\label{P70}
\end{eqnarray}
\begin{eqnarray}
\omega_{\delta \varphi_k}^2 & \equiv &
\frac{k^2}{a^2}+\frac{M^4}{\phi_c^2}
\left(\frac{\phi_0^2}{\phi_c^2}+1 \right)^
{\frac12 q-2} \left\{(q-1)
\frac{\phi_0^2}{\phi_c^2}+1 \right\} 
+\frac12 \langle\delta\phi^2\rangle
\frac{M^4}{\phi_c^4} \left(\frac{\phi_0^2}{\phi_c^2}+1 \right)^
{\frac12 q-4} \nonumber \\ 
& \times &
(q-2) \left\{ (q-1) (q-3)\frac{\phi_0^4}{\phi_c^4}
+6(q-3)\frac{\phi_0^2}{\phi_c^2}+3 \right\}+g^2\langle\chi^2
\rangle -\frac34\left(\frac{2\ddot{a}}{a}
+\frac{\dot{a}^2}{a^2}\right),
\label{P4}
\end{eqnarray}
\begin{eqnarray}
\omega_{X_k}^2 \equiv \frac{k^2}{a^2}+m_{\chi}^2
+g^2(\phi_0^2+\langle \delta \phi^2 \rangle)
-\frac34\left(\frac{2\ddot{a}}{a}
+\frac{\dot{a}^2}{a^2}\right),
\label{P5}
\end{eqnarray}
where we have considered the 
$\langle \delta \phi^2 \rangle^n$ terms up to $n=1$.
The higher order terms do not significantly contribute 
to the evolution of the system.
We find from Eq.~$(\ref{P70})$ that the frequency
of the $\varphi_0$ field changes even 
during one oscillation of the $\varphi_0$ field.
At the stage when the particles are not
sufficiently produced
($\langle \delta \phi^2 \rangle, \langle 
\chi^2 \rangle \ll \tilde{\phi}_0^2$), the second and
third terms in Eq.~$(\ref{P70})$ are negligible.
Moreover, since the last term in Eq.~$(\ref{P70})$
is estimated by using Eqs.~$(\ref{B16})$ and 
$(\ref{B15})$ as
\begin{eqnarray}
-\frac34\left(\frac{2\ddot{a}}{a}
+\frac{\dot{a}^2}{a^2}\right)
\approx \frac34 \kappa^2
\left[ \frac12 \dot{\phi_0}^2-V(\phi_0) \right],
\label{P30}
\end{eqnarray}
this term is also negligible compared with the first 
term in Eq.~$(\ref{P70})$ for the typical values of 
$\phi_c$ ($~\mbox{\raisebox{-1.ex}{$\stackrel
     {\textstyle<}{\textstyle \sim}$}}~10^{-3}M_{\rm pl}$).
Then the effective mass of the inflaton in the first stage 
of the oscillating inflation for  two limiting cases
$\phi_0^2 \gg \phi_c^2$ and $\phi_0^2 \ll \phi_c^2$ 
is given by
\begin{eqnarray}
m_{\phi}= \cases{
                \frac{M^2}{\phi_c} \left(\frac{\phi_c}{\phi_0}
                \right)^{1-\frac{q}{2}} & ($\phi_0^2 \gg \phi_c^2$)
                \cr
                \frac{M^2}{\phi_c} & ($\phi_0^2 \ll \phi_c^2$)
                \cr
                }
\label{P31}
\end{eqnarray}
Note that $m_{\phi}$ for $\phi_0^2 \gg \phi_c^2$
is smaller than $m_{\phi}$ for $\phi_0^2 \ll \phi_c^2$
because the potential is 
flat for $\phi_0^2 \gg \phi_c^2$.
As time passes and particles are produced, 
back reaction effects due to the increase of 
$\langle \delta\phi^2 \rangle$ and 
$\langle \chi^2 \rangle$ become relevant
and the coherent oscillation of the $\varphi_0$
field is prevented by the change of the effective mass
$m_{\phi}$.

In the frequency $(\ref{P4})$ of the $\delta \varphi_k$
field, since we consider the situation of $0<q<1$,
the second term becomes negative
in the region of $\phi_0^2~\mbox{\raisebox{-1.ex}{$\stackrel
     {\textstyle>}{\textstyle\sim}$}}~\phi_c^2$.
Moreover, in the small vicinity of 
$\phi_0^2~\mbox{\raisebox{-1.ex}{$\stackrel
     {\textstyle<}{\textstyle \sim}$}}~\phi_c^2$, this term rapidly grows and
takes large positive value $\sim M^4/\phi_c^2$.
These peculiar behaviors of the frequency lead to 
the enhancement of the $\phi$ particle during 
the stage of the oscillating inflation.

As for the $\chi$ field, 
since the oscillation of the $\phi_0$ field is different 
from that of the ordinary inflaton potential,
the increase of $\langle \chi^2\rangle$ occurs in a 
different way. 
In this case, although analytic investigations of the evolution of the 
$\chi$ particle are rather difficult, we shall examine the parameter range
of the critical value $\phi_c$ and the coupling constant $g$ 
where the $\chi$ particle production takes place efficiently.

With regard to the initial conditions of the fluctuation,
we choose the states which correspond to the conformal vacuum as
\begin{eqnarray}
\delta \varphi_k(0)=\frac{1}{\sqrt{2\omega_{\delta \varphi_k (0)}}},~~~
\delta \dot{\varphi}_k(0)=-i \omega_{\delta \varphi_k(0)} 
\delta \varphi_k(0),
\label{P7}
\end{eqnarray}
\begin{eqnarray}
X_k(0)=\frac{1}{\sqrt{2\omega_{X_k (0)}}},~~~
\dot{X}_k(0)=-i \omega_{X_k(0)} 
X_k(0),
\label{P8}
\end{eqnarray}
We investigate the evolution of the fluctuations of 
$\phi$ and $\chi$ fields with those initial conditions 
as the semiclassical problem.
In order to clarify the situation we consider, we first 
investigate the case when the coupling $g$ is zero
(i.e. one-field case)
and next investigate the case of $g \ne 0$ (two-field case).

\subsection{The case of $g=0$}   

In this subsection, we study the evolution of the inflaton quanta 
in the case of $g=0$.
This case was originally considered by Taruya\cite{Taruya}
including the metric perturbation.
However, since his results are obtained neglecting the back reaction 
effect of created particles, we investigate how this effect
would modify the growth of the fluctuation.
The first thing we should notice is that the strength of
the amplification of the fluctuation $\langle \delta\phi^2\rangle$
strongly depends on the critical value $\phi_c$. 
With the decrease of $\phi_c$, since the duration of the 
oscillating inflation becomes longer, parametric amplification
of the $\phi$ particle occurs more efficiently before the back reaction
effect becomes significant.

We investigate two concrete cases of $\phi_c=10^{-3}M_{\rm pl}$
and $\phi_c=10^{-4}M_{\rm pl}$  with the value of $q=0.1$ where 
the growth of the fluctuation can be expected.
When $q=0.1$, the initial value of $\phi_0$
is estimated as $\phi_0=1.4 \times 10^{-2} M_{\rm pl}$
by Eq.~$(\ref{B6})$.

Let us first consider the evolution of the $\phi_0$ field.
In the case of $\phi_c=10^{-3}M_{\rm pl}$, since the initial
value of $\phi_0$ is larger than $\phi_c$ only by one order of
magnitude, the number of $e$-foldings during the 
oscillating inflation is small [$N=0.93$ 
by Eq.~$(\ref{B4})$]. 
In Fig.~2, we depict the evolution of the $\phi_0$ field.
The amplitude $\tilde{\phi}_0$ gradually decreases due to 
the expansion of the universe,
and the oscillating inflation ends when 
$\tilde{\phi}_0$ is lowered to $\phi_c$.
Numerical calculations indicate that this occurs at
$\bar{t} \equiv M^2 t/M_{\rm pl} \approx 0.37$.
After this, the $\phi_0$ field oscillates around the core region 
$\phi_0^2~\mbox{\raisebox{-1.ex}{$\stackrel
     {\textstyle<}{\textstyle \sim}$}}~\phi_c^2$, whose bare mass is given by
$m_{\phi}=M^2/\phi_c$.
In the case of $\phi_c=10^{-4}M_{\rm pl}$, 
since the oscillating inflation continues until 
$\tilde{\phi}_0 \approx 10^{-4}M_{\rm pl}$,
we obtain the larger value of $e$-foldings $N=1.73$ compared with
the case of $\phi_c=10^{-3}M_{\rm pl}$.
Numerically, the oscillating inflation ceases 
at $\bar{t}\approx 0.45$ in this case.

With regard to the fluctuation of inflaton, the resonance
band is very broad and the growth rate of the fluctuation becomes 
of order unity. We can expect the enormous 
amplification of the wave modes especially within
the Hubble radius.
We show in Fig.~3 the evolution of the $\delta \phi_k$ field
for two cases of $\bar{k} \equiv k/(M^2/M_{\rm pl})=0.1$ and
$\bar{k}=100$.
We find that long-wave modes are not significantly enhanced
compared with the modes inside
the horizon scale. This result coincides with numerical 
calculations performed in Ref.~\cite{Taruya} which include 
the metric perturbation. As was presented in 
Ref.~\cite{Taruya}, the effect of the metric perturbation
appears in the term $2\kappa^2(V/H)^{\dot{}}$
of the equation of the Mukhanov variable
$Q \equiv \delta \phi -{\cal R} \dot{\phi}/H$, where
${\cal R}$ is the spatial curvature [See Eq.~(8) in
Ref.~\cite{Taruya}]. However, since this term decreases
faster than the term $V^{(2)}(\phi_0)$ in Eq.~$(\ref{B12})$,
we expect that adding this term to the equation 
of the fluctuation will not alter the evolution of the system.
We have numerically checked that the growth of the 
fluctuation in Fig.~3 is almost the same as in the case 
where the $2\kappa^2(V/H)^{\dot{}}$ term is included.
In the single-field case,
there exists an exact solution for the Mukhanov variable
in the long wavelength limit [See Eq.~(9) in
Ref.~\cite{Taruya}], and we can confirm that 
the amplitude of the fluctuation remains nearly constant.
On the other hand, since the instability band is 
broad in the present model even for the large $k$,
the momentum modes up to 
$k^2/a^2~\mbox{\raisebox{-1.ex}{$\stackrel
     {\textstyle<}{\textstyle \sim}$}}~M^4/{\phi_c}^2$ inside the Hubble horizon
are effectively enhanced.
In Fig.~3, we find that the fluctuation 
$\delta \phi_k$ rapidly grows at the stage of
the oscillating inflation for the case of $\bar{k}=100$. 
However, this increase is 
suppressed at $\bar{t}\approx 0.12$, before the oscillating 
inflation terminates at $\bar{t}\approx 0.37$.
This is due to the fact that the back reaction effect of 
created particles becomes significant.
We have numerically confirmed that the fluctuation
continues to grow during the oscillating inflation  
if we neglect the back reaction effect as in 
Ref.~\cite{Taruya}.
The increase of $\langle\delta\phi^2\rangle$
effectively changes both frequencies of
$\varphi_0$ and $\delta\varphi_k$ fields as is found by 
Eqs.~$(\ref{P70})$ and $(\ref{P4})$, and 
the growth of the fluctuation finally stops
when the $\langle\delta\phi^2\rangle$ terms become
comparable to the preceding terms in 
Eqs.~$(\ref{P70})$ and $(\ref{P4})$.
It is expected that the back reaction effect becomes 
significant when $\langle\delta\phi^2\rangle$ 
increases of order $\phi_c^2$. 
In the case of $\phi_c=10^{-3} M_{\rm pl}$,
numerical calculations show that 
the maximum value of the fluctuation is 
$\langle\delta\phi^2\rangle_f \approx 10^{-7} 
M_{\rm pl}^2$ at $\bar{t} \approx 0.12$
(See Fig.~4). 
In this case, parametric amplification of
the $\phi$ particle terminates
when $\langle\delta\phi^2\rangle$ increases up to 
$\langle\delta\phi^2\rangle_f \approx 0.1 \phi_c^2$. 
As is seen in Fig.~2, the coherent oscillation of the
$\phi_0$ field is hardly broken after $\bar{t} \approx 0.12$
where the variance of the $\phi$ particle reaches the 
maximum value. 
This means that the growth of $\langle\delta\phi^2\rangle$
affects the evolution of the $\delta\phi_k$ field 
more strongly than that of the $\phi_0$ field.

With the decrease of $\phi_c$, since the oscillating inflation 
continues longer, the growth rate of the fluctuation 
in the first stage becomes larger.
However, the final variance 
$\langle\delta\phi^2\rangle_f$ is suppressed as 
$\phi_c$ becomes smaller.
We depict in Fig.~5 the evolution of the 
fluctuation $\langle\delta\phi^2\rangle$ in the 
case of $\phi_c=10^{-4}M_{\rm pl}$.
In this case, although the oscillating inflation 
continues until $\bar{t} \approx 0.4$,
the growth of the fluctuation stops much earlier:
$\bar{t} \approx 0.05$.
Although the initial growth rate is larger than in the case 
of $\phi_c=10^{-3}M_{\rm pl}$, the 
production of the $\phi$ particle itself 
terminates the growth of the fluctuation.
This tendency is stronger with the decrease
of $\phi_c$, and the maximum fluctuation 
becomes smaller.
We have found the following relation
for the typical value of 
$\phi_c~\mbox{\raisebox{-1.ex}{$\stackrel
     {\textstyle<}{\textstyle \sim}$}}~
10^{-3}M_{\rm pl}$ as 
\begin{eqnarray}
\langle\delta\phi^2\rangle_f 
\approx 0.1 \phi_c^2.
\label{P9}
\end{eqnarray}
Although the $\phi$ particle production is 
possible at the initial stage of the oscillating inflation,
the final variance is strongly suppressed
with the decrease of $\phi_c$ as Eq.~$(\ref{P9})$.
After the amplitude of the inflaton field drops 
under $\tilde{\phi}_0~\mbox{\raisebox{-1.ex}{$\stackrel
     {\textstyle<}{\textstyle \sim}$}}~\phi_c$, 
the system enters the ordinary reheating stage
where the universe expands deceleratedly.
In this stage, the growth of 
the fluctuation $\langle\delta\phi^2\rangle$ 
can be no longer expected because the inflaton field
behaves as the massive 
inflaton whose mass is $m_{\phi} =M^2/\phi_c^2$.

In the next subsection, we consider another field $\chi$  
coupled to inflaton and analyze how $\chi$
particles are produced during the oscillating inflation
and subsequent oscillating phase by parametric resonance.

\subsection{The case of $g \ne 0$}   

Next, we consider the production of the $\chi$ particle 
coupled to the inflaton field.
In the ordinary picture of the slow-roll inflation,
the $\chi$ particle production is inefficient
during the inflationary phase.
In the present model, however, since the inflaton 
field oscillates rapidly during the oscillating inflation,
$\chi$ particles can be produced in this stage by
parametric resonance.

The strength of resonance depends on the coupling $g$, 
the amplitude of $\phi_0$
$(=\tilde{\phi}_0)$, and the mass of the inflaton field
$(=m_{\phi})$.
In the model of the scalar field $\chi$ coupled to the massive
inflaton $\phi$: $V(\phi, \chi)=m^2\phi^2/2+g^2\phi^2\chi^2/2$,
the equation of the $\chi$ field is reduced to the Mathieu
equation\cite{Mathieu} at the linear stage.
In this model, the resonance parameter
$q \equiv g^2\tilde{\phi}^2/(4m^2)$
is an important factor in determining whether the $\chi$
particle production is efficient or not\cite{KLS2}.
When $q$ is sufficiently large initially as $q_i \gg 1$,
parametric resonance turns on from broad resonance regimes.
Although $q$ decreases with the decrease of $\tilde{\phi}_0$
due to the cosmic expansion, the fluctuation
$\langle\chi^2\rangle$ grows quasiexponentially 
until $q$ drops under unity
or the back reaction effect of created particles becomes
significant.

In the oscillating inflation model, since the mass 
$m_{\phi}^2=\frac{M^4}{\phi_c^2}\left(
\frac{\phi_0^2}{\phi_c^2}+1\right)^{q/2-1}$
continually changes during one oscillation of inflaton,  
the $\varphi_0$ field does not oscillate sinusoidally and 
the method based on the Mathieu equation 
is not valid except when $\tilde{\phi}_0$
drops under $\phi_c$. 
In the frequency $(\ref{P5})$ of the $X_k$ field, 
since the last term is negligible when the $\chi$
particle production occurs sufficiently, the equation 
of the $\chi$ field for the case of 
$\langle\delta\phi^2\rangle \ll \tilde{\phi}_0^2$
is approximately written by
\begin{eqnarray}
\frac{d^2}{dz^2} X_k
+\left[\frac{\bar{k}^2}{a^2}+\bar{m}_{\chi}^2
+\frac{g^2 \phi_0^2}{m_{\phi}^2} \right] X_k
\approx 0,
\label{P12}
\end{eqnarray}
where $z \equiv m_{\phi}t$, $\bar{k}\equiv k/m_{\phi}$,
and $\bar{m}_{\chi}\equiv m_{\chi}/m_{\phi}$.
When the initial value of 
$g^2 \phi_0^2/m_{\phi}^2$ is large, 
parametric resonance works out more efficiently.
The initial value of $\phi_0$ estimated by 
Eq.~$(\ref{B6})$ for the typical value of $q$ 
is smaller than the initial 
value $\phi_0\sim 0.2$-$0.3 M_{\rm pl}$ in the massive
inflaton model. Moreover, as is found by
Eq.~$(\ref{B17})$,
the amplitude of the $\phi_0$ field decays faster than 
in the model of the massive inflaton
with $\tilde{\phi}_0 \propto t^{-1}$.

The effective mass of the $\phi_0$ field is approximately
written in two asymptotic regions as Eq.~$(\ref{P31})$, 
and depends on the critical value $\phi_c$.
The mass $m_{\phi}\approx M^2/\phi_c^2$ around 
$\phi_0^2~\mbox{\raisebox{-1.ex}{$\stackrel
     {\textstyle<}{\textstyle \sim}$}}~\phi_c^2$ plays the more important
role than that of $\phi_0^2~\mbox{\raisebox{-1.ex}{$\stackrel
     {\textstyle>}{\textstyle\sim}$}}~\phi_c^2$ 
for the development of the fluctuation, because 
$\chi$ particles are mainly produced
in the vicinity of $\phi_0=0$ where the $\phi_0$ field 
changes nonadiabatically.
During one oscillation of the $\phi_0$ field,
$m_{\phi}$ gradually gets larger as  $\phi_0$ 
approaches the minimum of the potential
for the fixed value of $\phi_c$.
Moreover, for the typical value of $\phi_c~\mbox{\raisebox{-1.ex}{$\stackrel
     {\textstyle<}{\textstyle \sim}$}}~10^{-3}
M_{\rm pl}$ where the oscillating inflation occurs, 
$m_{\phi}=M^2/\phi_c$ is greater than the mass
$\sim 10^{-6}M_{\rm pl}$ in the model
of the massive inflaton.
For example, in the case of 
$\phi_c=10^{-3}M_{\rm pl}$ and $q=0.1$,
$M^2/\phi_c=1.57 \times 10^{-4} M_{\rm pl}$;
for $\phi_c=10^{-4}M_{\rm pl}$ and $q=0.1$,
$M^2/\phi_c=1.40 \times 10^{-3} M_{\rm pl}$.

In addition to the fact that $\tilde{\phi}_0$ decreases 
faster than in the model of the massive inflaton,
the larger mass $m_{\phi}$ results in the restriction 
of the coupling constant $g$ for
an efficient particle production.
Since the dependence of $M^2$ for the value 
$\phi_c$ is weak, the mass $M^2/\phi_c$
monotonically increases with the decrease 
of $\phi_c$.
This means that parametric 
resonance does not take place unless we choose
a large coupling constant $g$
as $\phi_c$ decreases.
 When $\phi_c$ is fairly large as
 $\phi_c~\mbox{\raisebox{-1.ex}{$\stackrel
     {\textstyle>}{\textstyle\sim}$}}~0.1 M_{\rm pl}$, 
 $M^2/\phi_c$ becomes comparable to the mass
 of the massive inflaton model, but in this case
 the creation of $\phi$ particles can not be expected
 because of the absence of the oscillating 
 inflationary phase.
 What we are interested in is the case of 
 $\phi_c~\mbox{\raisebox{-1.ex}{$\stackrel
     {\textstyle<}{\textstyle \sim}$}}~10^{-3} M_{\rm pl}$ where the 
 oscillating inflation occurs and both of
 $\phi$ and $\chi$ particles are generated.
 In what follows, we examine the $\chi$ particle 
 production in two cases of 
 $\phi_c=10^{-3}M_{\rm pl}$
 and $\phi_c=10^{-4}M_{\rm pl}$ for the massless
 $\chi$ particle, and add the discussion of
 the case where the mass of the $\chi$ particle 
 is included at the final of this section.
 
When $\phi_c=10^{-3}M_{\rm pl}$,
numerical calculations indicate
that the increase of $\langle \chi^2 \rangle$ can take place 
for $g~\mbox{\raisebox{-1.ex}{$\stackrel
     {\textstyle>}{\textstyle\sim}$}}~5 \times 10^{-3}$. However, parametric
resonance is weak for the case of $g~\mbox{\raisebox{-1.ex}{$\stackrel
     {\textstyle<}{\textstyle \sim}$}}~0.03$. 
As compared with the model of
 $V(\phi, \chi)=m^2\phi^2/2+g^2\phi^2\chi^2/2$ in which
the $\chi$ particle production takes place
for $g~\mbox{\raisebox{-1.ex}{$\stackrel
     {\textstyle>}{\textstyle\sim}$}}~10^{-4}$\cite{KLS2}, larger values 
of $g$ are required 
even in the case of $\phi_c=10^{-3}M_{\rm pl}$ 
where the oscillating inflation marginally occurs.
Since the fluctuation $\langle \chi^2 \rangle$ does not 
grow well for $g~\mbox{\raisebox{-1.ex}{$\stackrel
     {\textstyle<}{\textstyle \sim}$}}~0.03$, the evolution
of the $\phi_0$ field is hardly affected by the $\chi$
particle production.
For the case of $g~\mbox{\raisebox{-1.ex}{$\stackrel
     {\textstyle>}{\textstyle\sim}$}}~0.05$, the back 
reaction effect of the created $\chi$ particle begins to
be significant.
We depict in Fig.~6 the evolution of $\langle\chi^2\rangle$
for two cases of $g=0.03$ and $g=0.07$.
Although parametric resonance evidently occurs
for $g=0.03$, 
this is rather inefficient and $\langle\chi^2\rangle$
takes the maximum value 
$\langle\chi^2\rangle_f=2.3 \times
10^{-11}M_{\rm pl}^2$ at $\bar{t}=0.177$.
In this case, the $\chi$ field deviates from
the resonance bands with the decrease of $\tilde{\phi}_0$
due to the cosmic expansion before the oscillating inflation
ceases at $\bar{t} \approx 0.37$.
Since the maximum fluctuation of 
$\langle\delta\phi^2\rangle$ is almost the same as the $g=0$
case, the back reaction effect of $\phi$ particles onto the
$\phi_0$ field can be marginally negligible.
Hence in the case of $g~\mbox{\raisebox{-1.ex}{$\stackrel
     {\textstyle<}{\textstyle \sim}$}}~0.03$, 
the coherent oscillation of the $\phi_0$ field is hardly 
prevented by the back reaction effects of both $\phi$
and $\chi$ particles.
When $g=0.07$, the evolution of the system
shows different characteristics compared with the case of
$g~\mbox{\raisebox{-1.ex}{$\stackrel
     {\textstyle<}{\textstyle \sim}$}}~0.03$. In Fig.~7, the evolution of the 
$\phi_0$ field for the $g=0.07$ case is depicted.
We find that the coherent oscillation of the $\phi_0$ field  
is broken for $\bar{t}~\mbox{\raisebox{-1.ex}{$\stackrel
     {\textstyle>}{\textstyle\sim}$}}~0.12$, which is different 
from the $g=0$ case in Fig.~2. 
As for the $\phi$ particle, the maximum fluctuation is 
$\langle\delta\phi^2\rangle_f=2.7\times 10^{-6}
M_{\rm pl}^2$ at $\bar{t}=0.12$.
This maximum variance is greater than in the case of $g=0$
by one order of magnitude (See Fig.~8).
As was noticed in the previous subsection, the growth
of  $\langle\delta\phi^2\rangle$ stops when the contribution
of the third term in the frequency (3.5) becomes comparable
to the second term.
Since the third term mostly takes negative values,
the increase of the $g^2\langle\chi^2\rangle$ term
due to the $\chi$ particle production generally assists
the $\phi$ particle production to continue longer.
This results in the larger value of the final fluctuation
$\langle\delta\phi^2\rangle$, and the back reaction effect
due to the $\phi$ particle production becomes relevant
at $\bar{t} \approx 0.12$. This behavior is found in 
Fig.~7.
However, since the $\phi_0$ field still 
oscillates after the back reaction effect of the $\phi$ 
particle becomes important, the $\chi$ particle continues
to be enhanced after $\bar{t} \approx 0.12$.
The fluctuation of $\langle\chi^2\rangle$ does not 
increase significantly in the initial stage of the 
oscillating inflation, because the expansion rate of 
the universe is large. 
$\langle\chi^2\rangle$ begins to increase rapidly
after $\bar{t}\approx 0.15$, and reaches the maximum
value $\langle\chi^2\rangle_f=2.4\times 10^{-7}
M_{\rm pl}^2$ at $\bar{t}=0.28$.
 The back reaction effect of the 
$\chi$ particles as well as the decrease of
$\tilde{\phi}_0$ terminates the parametric
amplification of $\chi$ particles.
When $g=0.1$, the increase of $\langle\chi^2\rangle$
is more significant and the final variance is
$\langle\chi^2\rangle_f=3.3 \times 10^{-6}
M_{\rm pl}^2$ .
In this case, since the maximum variance is larger than
in the case of $g=0.07$ by one
order of magnitude, the coherent oscillation of 
the $\phi_0$ field is more strongly prevented
by the growth of $\langle\chi^2\rangle$.
When $g~\mbox{\raisebox{-1.ex}{$\stackrel
     {\textstyle>}{\textstyle\sim}$}}~0.3$, the final variance is suppressed
because the back reaction effect becomes
quite significant.
We find that $\langle\chi^2\rangle_f$ takes the 
maximum value $\langle\chi^2\rangle_{\rm max}
\approx 10^{-5}M_{\rm pl}^2$ for $g \approx 0.3$
(See Fig.~9).

As for the momentum modes of the produced $\chi$
particle, we show the evolution of the $\chi_k$ field
in two cases of $\bar{k}=0$ and $\bar{k}=100$
for  $g=0.1$ in Fig.~10.
Since the larger $k$ makes the $\chi_k$ field deviate from 
the resonance band in Eq.~$(\ref{P12})$,
higher momentum modes
of the $\chi$ particle are not sufficiently produced.
The low momentum modes mainly contribute to 
the growth of the fluctuation $\langle\chi^2\rangle$.
This property is the same as in the model of 
$V(\phi, \chi)=m^2\phi^2/2+g^2\phi^2\chi^2/2$\cite{KLS2}.
As was found in Ref.~\cite{mpre2}, there is a possibility
that long wavelength modes of the 
metric perturbation are amplified considering
the perturbed metric.
This issue will be discussed elsewhere\cite{comment2}.
 
In the case of $\phi_c=10^{-4}M_{\rm pl}$, we need 
further large values of $g$ to yield an efficient resonance.
Numerically, we find that the coupling is
required $g~\mbox{\raisebox{-1.ex}{$\stackrel
     {\textstyle>}{\textstyle\sim}$}}~0.01$ for the development
of the $\chi$ fluctuation.
Even when $g=0.1$, parametric resonance 
is not so efficient  compared with 
the case of $\phi_c=10^{-3}M_{\rm pl}$.
We depict in Fig.~11 the evolution of 
$\langle\chi^2\rangle$ for $g=0.07, 0.1, 0.5$ cases 
respectively.
When $g=0.07$, the maximum value of the
fluctuation is $\langle\chi^2\rangle_f=
1.2\times 10^{-10}M_{\rm pl}^2$,
which is much smaller than in the case of 
$\phi_c=10^{-3}M_{\rm pl}$ with the same value of $g$.
As the coupling increases further, we obtain larger 
values of the maximum fluctuation $\langle\chi^2\rangle_f$.
However, for $g~\mbox{\raisebox{-1.ex}{$\stackrel
     {\textstyle>}{\textstyle\sim}$}}~0.5$,
the back reaction effect of created $\chi$ particles becomes
significant and the final variance is suppressed.
In the case of $\phi_c=10^{-4}M_{\rm pl}$ and $g=0$, 
the fluctuation of the $\phi$ field rapidly grows
in the initial stage and reaches the maximum value 
$\langle\delta\phi^2\rangle_f \approx 10^{-9}M_{\rm pl}^2$ at
$\bar{t}=0.05$. Even if we take into account the interaction
with the $\chi$ field, this maximum variance is 
hardly altered because the increase of 
$\langle\chi^2\rangle$ is weak for
$\bar{t}~\mbox{\raisebox{-1.ex}{$\stackrel
     {\textstyle<}{\textstyle \sim}$}}~0.05$ as is found in Fig.~11.
The back reaction effect of $\phi$ particles onto
the $\phi_0$ field is marginally negligible
as the case of $g=0$.
When $g=0.1$, the final variance is 
$\langle\chi^2\rangle_f=2.5\times
10^{-9} M_{\rm pl}^2$, and the growth of the
fluctuation hardly affects the evolution of the 
$\phi_0$ field.
However, in the case of $g=0.5$, 
$\langle\chi^2\rangle_f$ increases up to
$\langle\chi^2\rangle_f=1.3\times
10^{-6} M_{\rm pl}^2$ at $\bar{t}\approx 0.17$.
In Fig.~12, we find that the increase of $\chi$
particles prevents the coherent
oscillation of  the $\phi_0$ field for 
$\bar{t}~\mbox{\raisebox{-1.ex}{$\stackrel
     {\textstyle>}{\textstyle\sim}$}}~0.17$.
When $g~\mbox{\raisebox{-1.ex}{$\stackrel
     {\textstyle>}{\textstyle\sim}$}}~0.5$, since the $\chi$
particle production stops by the back reaction 
effect, the final variance gradually decreases 
with the increase of $g$.
As a result, the maximum variance for the case of 
$\phi_c =10^{-4}M_{\rm pl}$  is 
$\langle\chi^2\rangle_{\rm max} \approx
10^{-6} M_{\rm pl}^2$ for $g \approx 0.5$.

With the decrease of $\phi_c$
($~\mbox{\raisebox{-1.ex}{$\stackrel
     {\textstyle<}{\textstyle \sim}$}}~10^{-5}M_{\rm pl})$, the development of the 
$\chi$ fluctuation is not expected unless $g$ is
unnaturally large.
Moreover, in this case, although the $\phi$ particle
production is possible in the initial stage of the 
oscillating inflation, the final fluctuation 
$\langle\delta\phi^2\rangle_f$ is strongly suppressed.
This means that it is difficult to obtain the sufficient
amount of $\phi$ and $\chi$ particles with the natural 
coupling $g~\mbox{\raisebox{-1.ex}{$\stackrel
     {\textstyle<}{\textstyle \sim}$}}~1$ in the case of 
$\phi_c~\mbox{\raisebox{-1.ex}{$\stackrel
     {\textstyle<}{\textstyle \sim}$}}~10^{-5}M_{\rm pl}$.
For large critical values $\phi_c>10^{-3}M_{\rm pl}$,
the $\chi$ particle
production occurs effectively for the natural coupling.
With the increase of $\phi_c$, the structure of resonance
for the $\chi$ field approaches that of the model 
$V(\phi,\chi)=m^2\phi^2/2+g^2\phi^2\chi^2/2$,
and the investigation based on the Mathieu equation
performed in Ref.~\cite{KLS2} can be applied.
Since the mass $m_{\phi}$ 
around the minimum of the inflaton potential
becomes smaller, the $\chi$ particle production 
occurs efficiently.
In this case, however, since the duration of the oscillating 
is too short, the $\phi$ particle production is hardly
expected.
In the case of the massless $\chi$ particle, we conclude that 
both $\phi$ and $\chi$ particles are most efficiently created
in the parameter range $\phi_c=10^{-3}$-$10^{-4}M_{\rm pl}$ for the natural
coupling $g$.

Finally, we discuss the case where the mass 
of the $\chi$ particle is included.
We expect that the large mass
makes the $\chi$ field deviate from the resonance band 
by the relation $(\ref{P12})$.
Let us consider the case of $\phi_c=10^{-3}M_{\rm pl}$
and $q=0.1$.
In the massless case, the $\chi$ particle production
occurs relevantly  for $g~\mbox{\raisebox{-1.ex}{$\stackrel
     {\textstyle>}{\textstyle\sim}$}}~0.05$.
If the mass of the $\chi$ particle is taken into account,
numerical calculations show that heavy particles up
to the order of  $m_{\chi}=10^{14}$ GeV 
can be produced enough for 
$g~\mbox{\raisebox{-1.ex}{$\stackrel
     {\textstyle>}{\textstyle\sim}$}}~0.05$.
In Fig.~13, the evolution of the fluctuation 
$\langle\chi^2\rangle$ is depicted in two cases of 
$m_{\chi}=3 \times 10^{14}$ GeV and 
$m_{\chi}=1 \times 10^{15}$ GeV for $g=0.05$.
The final variance 
$\langle\chi^2\rangle_f \approx 10^{-8}M_{\rm pl}^2$
with mass $m_{\chi}=3 \times 10^{14}$ GeV
is almost the same as in the case of the massless 
$\chi$ particle with the same coupling $g$.
This means that the massive $\chi$ particle whose mass is 
$m_\chi~\mbox{\raisebox{-1.ex}{$\stackrel
     {\textstyle<}{\textstyle \sim}$}}~3 \times 10^{14}$ GeV is produced 
with almost the same amount as  the massless 
$\chi$ particle for $g=0.05$.
However, the achieved amount of the produced $\chi$
particle decreases with the increase of $m_{\chi}$
for $m_\chi~\mbox{\raisebox{-1.ex}{$\stackrel
     {\textstyle>}{\textstyle\sim}$}}~3 \times 10^{14}$ GeV.
In the case of $g=0.05$, although the $\chi$ particle 
with mass $m_\chi~\mbox{\raisebox{-1.ex}{$\stackrel
{\textstyle<}{\textstyle \sim}$}}~7 \times 10^{14}$ GeV 
can be created,
it is difficult to generate the massive particle whose mass
is $m_{\chi}~\mbox{\raisebox{-1.ex}{$\stackrel
     {\textstyle>}{\textstyle\sim}$}}~1 \times 10^{15}$
GeV (See Fig.~13).
In the model of $V(\phi,\chi)=
m^2\phi^2/2+g^2\phi^2\chi^2/2$, analytic investigations
show that the $\chi$ particle which satisfies the condition
$m_{\chi}~\mbox{\raisebox{-1.ex}{$\stackrel
     {\textstyle<}{\textstyle \sim}$}}~\sqrt{g}~10^{15}$ GeV can be 
generated\cite{KLS1,baryogenesis}.
When $m_{\chi}=10^{14}$ GeV, numerical calculations
performed in Ref.~\cite{KT1} reveal that the $\chi$ particle
production occurs for $g~\mbox{\raisebox{-1.ex}{$\stackrel
     {\textstyle>}{\textstyle\sim}$}}~0.06$\cite{comment3}.
This condition is almost the same as in the oscillating inflation
model in spite of the difficulty in producing the {\it massless}
$\chi$ particle without the larger $g$ compared with the $V(\phi,\chi)=
m^2\phi^2/2+g^2\phi^2\chi^2/2$ model.
The reason is simple.
In Eq.~$(\ref{P12})$, the massive $\chi$ particle production
occurs when the initial value of the resonance term 
$g^2\phi_0^2/m_{\phi}^2$ is much greater than unity 
[$(g^2\phi_0^2/m_{\phi}^2)_i \gg 1$] and $\bar{m}_{\chi}^2$
is much smaller than this term $[\bar{m}_{\chi}^2 \ll
(g^2\phi_0^2/m_{\phi}^2)_i]$.
In the case of $\phi_c=10^{-3}M_{\rm pl}$, $q=0.1$, and
$g=0.05$, since $M =3.96 \times 10^{-4}M_{\rm pl}$
by Eq.~$(\ref{B22})$, we obtain
$(g^2\phi_0^2/m_{\phi}^2)_i \approx 4\times 10^3$.
Although $\tilde{\phi}_0$ decreases by the cosmic expansion
and $m_{\phi}$ becomes large in the vicinity of $\phi_0=0$,
the massive $\chi$ particle production is possible as long as
$g^2\tilde{\phi}_0^2/m_{\phi}^2 \gg 1$ and 
$\bar{m}_{\chi}^2 \ll g^2\tilde{\phi}_0^2/m_{\phi}^2$.
Since the mass $m_{\phi}$ in this case is estimated as
$m_{\phi}\approx 10^{-5}M_{\rm pl}$ for 
$\phi_0^2 \gg \phi_c^2$
and $m_{\phi}\approx 10^{-4}M_{\rm pl}$ for 
$\phi_0^2 \ll \phi_c^2$ by Eq.~$(\ref{P31})$, this is 
heavier than the mass $m \approx 10^{-6}M_{\rm pl}$
in the $V(\phi,\chi)=m^2\phi^2/2+g^2\phi^2\chi^2/2$ model.
Even in the case where the normalized mass is 
$\bar{m}_{\chi}=1$, this corresponds to the rather heavy
particle whose mass is 
$m_{\chi}~\mbox{\raisebox{-1.ex}{$\stackrel
     {\textstyle>}{\textstyle\sim}$}}~10^{14}$ GeV.
For the typical value $\phi_c~\mbox{\raisebox{-1.ex}{$\stackrel
  {\textstyle<}{\textstyle \sim}$}}~10^{-3}M_{\rm pl}$
in the oscillating inflation model, since $m_{\phi}$
is at least by one order larger than 
$m=10^{-6}M_{\rm pl}$, the massive $\chi$ particle whose 
mass is of order $m_{\chi}=10^{14}$ GeV can be generated
for the coupling $g$ where the massless $\chi$
production occurs. 
As the coupling $g$ increases, more massive particles are 
generated.  In the case of $\phi_c=10^{-3}M_{\rm pl}$,
we find that the production of the massive particle
whose mass is of order $m_{\chi}=10^{15}$ GeV
relevantly occurs
for the coupling $g~\mbox{\raisebox{-1.ex}{$\stackrel
     {\textstyle>}{\textstyle\sim}$}}~0.2$.
Even the GUT scale bosons $m_{\chi}=10^{16}$ GeV 
are produced in the initial stage of the oscillating inflation 
for the case of $g=1$, although the achieved amount
is small (See Fig.~14).

With the decrease of $\phi_c$, the larger values of $g$ 
are required in order to produce the massive 
$\chi$ particle
as is the same with the massless $\chi$ particle.
However, in the parameter range of $g$ where 
the massless $\chi$ particles are sufficiently produced, 
we can also expect the generation of 
the superheavy particles needed for the success of 
the GUT baryogenesis. For example, in the case of 
$\phi_c=10^{-4} M_{\rm pl}$,
numerical calculations indicate that 
$\chi$ particles up to the mass 
$m_{\chi}~\mbox{\raisebox{-1.ex}{$\stackrel
     {\textstyle<}{\textstyle \sim}$}}~10^{14}$ GeV and 
$m_{\chi}~\mbox{\raisebox{-1.ex}{$\stackrel
     {\textstyle<}{\textstyle \sim}$}}~10^{15}$ GeV are created for 
the coupling $g~\mbox{\raisebox{-1.ex}{$\stackrel
     {\textstyle>}{\textstyle\sim}$}}~0.07$ and 
     $g~\mbox{\raisebox{-1.ex}{$\stackrel
     {\textstyle>}{\textstyle\sim}$}}~0.3$ respectively.
It will be interesting to investigate how these 
superheavy bosons produced in the oscillating inflation 
model would affect the scenario of the GUT baryogenesis. 

\section{Concluding remarks and Discussions}    %
In this paper we have considered parametric
amplification of a scalar field $\chi$ coupled to an 
inflaton field $\phi$ as well as the enhancement 
of the fluctuation of the $\phi$ field
in the oscillating inflation model.
Since the inflaton potential $V(\phi)$ is
the non-convex type where $d^2 V/d\phi^2$ 
takes negative values in the regions not too far from 
the minimum of the potential, inflation is realized even 
during the oscillation of the inflaton field.
The oscillating inflation model is characterized
by the critical value of $\phi_c$ which indicates the scale
of the core part of the potential.
Inflation takes place while the $\phi$
field moves in the regions of 
$|\phi|~\mbox{\raisebox{-1.ex}{$\stackrel
     {\textstyle>}{\textstyle\sim}$}}~\phi_c$.
With the decrease of $\phi_c$, the total amount of inflation
becomes larger, because the duration during which the 
oscillating inflation takes place becomes longer.
However, it is rather difficult to obtain 
the sufficient number of
$e$-foldings such as $N~\mbox{\raisebox{-1.ex}{$\stackrel
     {\textstyle>}{\textstyle\sim}$}}~60$ 
only by the stage of the oscillating inflation, and 
the main contribution to the number of $e$-foldings is 
achieved by the ordinary slow-roll inflation 
which precedes the oscillating inflation. 
We have estimated the energy scale of the
oscillating inflation $M^4$ by fitting the primordial 
density perturbation observed by COBE.
The oscillating inflation terminates when the amplitude
of the $\phi_0$ field decreases to of the order $\phi_c$
by the expansion of the universe.

In this model, the fluctuation of the $\phi$ field
grows during the oscillating inflation 
due to the existence of the
non-convex part of the potential. 
We examined the growth of the fluctuation for
two typical cases $\phi_c=10^{-3}M_{\rm pl}$ and
$\phi_c=10^{-4}M_{\rm pl}$ including the back reaction
effect of created particles based on the
Hartree approximation.
We have found that the wave modes 
inside the Hubble horizon are effectively
enhanced, while the growth of long-wave modes 
is small as was shown in Ref.~\cite{Taruya}.
Although the growth rate of the fluctuation 
$\langle\delta\phi^2\rangle$ in the initial
stage of the oscillating inflation becomes
larger with the decrease of $\phi_c$, 
the final fluctuation is significantly suppressed.
This indicates that the generation of the $\phi$ particle
itself plays a crucial role for the termination of 
parametric resonance.
As for the final variance of the $\phi$ particle, we find 
the relation $\langle\delta\phi^2\rangle_f \approx 0.1
\phi_c^2$ for the case of 
$\phi_c~\mbox{\raisebox{-1.ex}{$\stackrel
     {\textstyle<}{\textstyle \sim}$}}~10^{-3}M_{\rm pl}$.

With regard to the $\chi$ field coupled to inflaton, the enhancement 
of the fluctuation mainly occurs when the $\phi_0$ field moves 
nonadiabatically around the minimum of the potential.  Since the mass 
$m_{\phi}$ for $|\phi_0|~\mbox{\raisebox{-1.ex}{$\stackrel
     {\textstyle<}{\textstyle \sim}$}}~\phi_c$ is approximately written as $m_{\phi} 
\approx M^2/\phi_c$, it is larger than that of the massive inflaton of the 
chaotic inflation model for the value of 
$\phi_c~\mbox{\raisebox{-1.ex}{$\stackrel
     {\textstyle<}{\textstyle \sim}$}}~10^{-3}M_{\rm pl}$.  As $m_{\phi}$ becomes larger, this 
restricts the effective $\chi$ particle production.  Moreover,  the 
amplitude of the homogeneous inflaton field decreases faster than in the 
model of the massive inflaton.  This means that the $\chi$ particle 
production is not possible unless we select rather large values of the 
coupling constant $g$.  This tendency becomes stronger with the decrease of 
$\phi_c$.  For example, in the case of $\phi_c=10^{-3}M_{\rm pl}$ and 
$q=0.1$, the coupling $g~\mbox{\raisebox{-1.ex}{$\stackrel
     {\textstyle>}{\textstyle\sim}$}}~0.05$ is required to yield the sufficient 
$\chi$ particle production, the value of which is two orders of magnitude larger 
than in the model of $V(\phi, \chi)=m^2\phi^2/2+ g^2\phi^2\chi^2/2$.  For 
$g~\mbox{\raisebox{-1.ex}{$\stackrel
     {\textstyle>}{\textstyle\sim}$}}~0.05$, the back reaction effect onto the $\phi_0$ field due to the 
$\chi$ particle production is significant, and finally terminates 
parametric resonance.  The final variance is suppressed by the back 
reaction effect with the increase of $g$, and the maximal variance for the 
$\phi_c=10^{-3}M_{\rm pl}$ case is $\langle\chi^2\rangle_f \approx 
10^{-5}M_{\rm pl}^2$ when $g \approx 0.3$.  In the case of 
$\phi_c=10^{-4}M_{\rm pl}$, we need $g~\mbox{\raisebox{-1.ex}{$\stackrel
     {\textstyle>}{\textstyle\sim}$}}~0.1$ for an effective 
resonance, although the $\chi$ particle production occurs for 
$g~\mbox{\raisebox{-1.ex}{$\stackrel
     {\textstyle>}{\textstyle\sim}$}}~0.01$. The maximal variance for the 
$\phi_c=10^{-4}M_{\rm pl}$ case is $\langle\chi^2\rangle_f \approx 
10^{-6}M_{\rm pl}^2$ when $g \approx 0.5$.  As $\phi_c$ decreases further, 
we no longer expect the growth of the $\chi$ fluctuation 
with the natural coupling $g~\mbox{\raisebox{-1.ex}{$\stackrel
     {\textstyle<}{\textstyle \sim}$}}~1$.
Moreover, since the final fluctuation of the $\phi$ particle 
is also suppressed, the small $\phi_c$ such as 
$\phi_c~\mbox{\raisebox{-1.ex}{$\stackrel
     {\textstyle<}{\textstyle \sim}$}}~10^{-5}M_{\rm pl}$ is not favorable 
for the production of a sufficient amount of $\phi$ and 
$\chi$ particles.

In the oscillating inflation model, 
the superheavy $\chi$ particle with mass greater than 
$m_{\chi}=10^{14}$ GeV can be copiously produced.
For example, in the case of $\phi_c=10^{-3}
M_{\rm pl}$, we find that 
the $\chi$ particle whose mass is 
of order $10^{14}$GeV and $10^{15}$GeV is 
effectively enhanced for $g~\mbox{\raisebox{-1.ex}{$\stackrel
     {\textstyle>}{\textstyle\sim}$}}~0.05$ and 
$g~\mbox{\raisebox{-1.ex}{$\stackrel
     {\textstyle>}{\textstyle\sim}$}}~0.2$ respectively.
The GUT scale boson $m_{\chi}\sim
10^{16}$ GeV can be also generated in the initial stage 
if the coupling $g$ is of order unity, although the final
amount is small.
In Ref.~\cite{baryogenesis}, the authors considered
the GUT scale baryogenesis with the massive inflaton
assuming that some of the initial inflaton energy is
efficiently transferred to the bosons with the mass
$10^{14}$ GeV. These massive bosons decay into lighter
particles, and produce the net baryon number.
Although we do not consider such a decaying process in
this paper, it is of interest how the 
production of the superheavy particle 
would affect the baryon asymmetry in the universe.

In this paper we make use of the Hartree approximation,
which is essentially the mean field approximation.
This does not include the rescattering effect which becomes 
important as $\chi$ particles are sufficiently produced.
The scattering between $\phi$ and $\chi$ particles 
may reduce the final amount of fluctuations.
For a complete study including the nonlinear effect of
the particle production, we should perform the lattice
simulations and compare them with the mean 
field approximation performed in this paper.

Finally, we comment on the case where the metric 
perturbation is included. In the single field case, 
even if we consider the effect of the metric perturbation, 
the $V^{(2)}(\phi_0)$ term  in Eq.~$(\ref{B12})$
is much more important than the 
$2\kappa^2(V/H)^{\dot{}}$ term which is the 
gravitational origin. 
Hence the situation is almost the same as the case 
when the metric perturbation is neglected.
As we have showed, the long-wave
modes are not significantly enhanced compared 
with the modes inside the Hubble Horizon
in the single-field case. 
This result is consistent with other inflation models in the single-field
case\cite{mpre1}.
In the two-field case, it was suggested that super-Hubble 
metric perturbations can be amplified in broad classes 
of models\cite{mpre2}. 
In the present model, since the $\chi$ fluctuation of 
super-Hubble modes will be exponentially suppressed
for large values of $q \gg 1$\cite{mpre3}, 
we may not expect the enhancement of  super-Hubble metric 
perturbations significantly.
However, it is worth investigating to investigate these issues
including mode-mode coupling between field and metric 
perturbations, because this may lead to some imprints
on the spectrum of density perturbations.
These issues are under consideration.

\section*{ACKOWLEDGEMENTS}
The author would like to thank  Bruce A. Bassett, Kei-ichi Maeda, Atsushi Taruya, 
Takashi Torii, and Hiroki Yajima for useful discussions.
This work was supported partially by a Grant-in-Aid for  Scientific
Research Fund of the Ministry of Education, Science and Culture
(No. 09410217 and Specially Promoted Research No. 08102010),
and by the Waseda University Grant for 
Special Research Projects.


\newpage
\vskip 2cm
\begin{flushleft}
{Figure Captions}
\end{flushleft}
\noindent
\parbox[t]{2cm}{FIG. 1:\\~}\ \
\parbox[t]{8cm}
{The potential for $q=0.1$ in the oscillating inflation model,
which is composed of the non-convex region $|\phi|~\mbox{\raisebox{-1.ex}{$\stackrel
     {\textstyle>}{\textstyle\sim}$}}~\phi_c$ 
and the core region $|\phi|~\mbox{\raisebox{-1.ex}{$\stackrel
     {\textstyle<}{\textstyle \sim}$}}~\phi_c$.
}\\[1em]
\noindent
\parbox[t]{2cm}{FIG. 2:\\~}\ \
\parbox[t]{8cm}
{The evolution of the $\phi_0$ field in the case of
$\phi_c=10^{-3}M_{\rm pl}$, $q=0.1$, and $g=0$.
The stage of the oscillating inflation starts from 
$\phi_0 \approx 1.4 \times 10^{-2} M_{\rm pl}$,
and ends at $\bar{t}\approx 0.37$ when the amplitude
of the $\phi_0$ field becomes of order $\phi_c$.
 }\\[1em]
\noindent
\parbox[t]{2cm}{FIG. 3:\\~}\ \
\parbox[t]{8cm}
{The evolution of the real part of the fluctuation 
$\delta\phi_k$ for two momentum modes of
$\bar{k}=0.1$ and $\bar{k}=100$
in the case of $\phi_c=10^{-3}M_{\rm pl}$,
$q=0.1$, and $g=0$.
For the long wave mode $\bar{k}=0.1$, 
the growth of the fluctuation is weak,
but for the mode $\bar{k}=100$, the fluctuation 
grows rapidly until
the back reaction effect becomes significant at 
$\bar{t}\approx 0.12$.
}\\[1em]
\noindent
\parbox[t]{2cm}{FIG. 4:\\~}\ \
\parbox[t]{8cm}
{The evolution of $\langle \delta \phi^2\rangle$
 in the case of $\phi_c=10^{-3}M_{\rm pl}$,
$q=0.1$, and $g=0$.
The fluctuation reaches the maximum value
$\langle \delta \phi^2\rangle_f \approx 10^{-7}
M_{\rm pl}^2$ at $\bar{t} \approx 0.12$. 
}\\[1em]
\noindent
\parbox[t]{2cm}{FIG. 5:\\~}\ \
\parbox[t]{8cm}
{The evolution of $\langle \delta \phi^2\rangle$
 in the case of $\phi_c=10^{-4}M_{\rm pl}$,
 $q=0.1$, and $g=0$.
 Although the initial growth rate is larger
 than in the case of $\phi_c=10^{-3}M_{\rm pl}$,
 the fluctuation soon reaches the maximum value 
 $\langle \delta \phi^2\rangle_f \approx 8.0
 \times 10^{-10} M_{\rm pl}^2$,
 which is by two orders smaller
 than in the case of $\phi_c=10^{-3}M_{\rm pl}$.
}\\[1em]
\noindent
\parbox[t]{2cm}{FIG. 6:\\~}\ \
\parbox[t]{8cm}
{The evolution of the fluctuation 
$\langle \chi^2\rangle$ 
in the case of $\phi_c=10^{-3}M_{\rm pl}$ and $q=0.1$
for $g=0.03$ (bottom) and  $g=0.07$ (top).
For $g=0.03$, the maximum value of the fluctuation
is $\langle\chi^2\rangle_f=2.3 \times
10^{-11}M_{\rm pl}^2$; and for $g=0.07$,
$\langle\chi^2\rangle_f=2.4 \times 
10^{-7}M_{\rm pl}^2$.
}\\[1em]
\noindent
\parbox[t]{2cm}{FIG. 7:\\~}\ \
\parbox[t]{8cm}
{The evolution of the $\phi_0$ field in the case of 
$\phi_c=10^{-3}M_{\rm pl}$, $q=0.1$, and $g=0.07$.
The coherent oscillation is a bit broken at $\bar{t}\approx 0.12$
due to the increase of $\langle\delta\phi^2\rangle$.
With the increase of  $\langle\chi^2\rangle$, this also
changes the frequency of the $\phi_0$ field at 
$\bar{t} \approx 0.28$.
}\\[1em]
\noindent
\parbox[t]{2cm}{FIG. 8:\\~}\ \
\parbox[t]{8cm}
{The evolution of $\langle\delta\phi^2\rangle$ in the
case of  $\phi_c=10^{-3}M_{\rm pl}$, $q=0.1$,
and $g=0.07$.
With the growth of $\langle\chi^2\rangle$, this assists
the $\phi$ particle production, and the final variance 
$\langle\delta\phi^2\rangle_f=2.7\times
10^{-6}M_{\rm pl}^2$ becomes larger than in the case of 
$\phi_c=10^{-3}M_{\rm pl}$ and $g=0$ by one order of
magnitude.
}\\[1em]
\noindent
\parbox[t]{2cm}{FIG. 9:\\~}\ \
\parbox[t]{8cm}
{The final variance $\langle\chi^2\rangle_f$ 
as the function of $g$ in the case of 
$\phi_c=10^{-3}M_{\rm pl}$ and $q=0.1$.
We find that $\langle\chi^2\rangle_f$ 
takes the maximum value 
$\langle\chi^2\rangle_{\rm max} \approx
10^{-5}M_{\rm pl}^2$ when $g \approx 0.3$.
}\\[1em]
\noindent
\parbox[t]{2cm}{FIG. 10:\\~}\ \
\parbox[t]{8cm}
{The evolution of the real part of the fluctuation 
$\chi_k$ for two momentum modes of
$\bar{k}=0$ and $\bar{k}=100$
in the case of $\phi_c=10^{-3}M_{\rm pl}$,
$q=0.1$, and $g=0.1$.
The long wave mode $\bar{k}=0$ is strongly
enhanced compared with the wave mode 
$\bar{k}=100$.
}\\[1em]
\noindent
\parbox[t]{2cm}{FIG. 11:\\~}\ \
\parbox[t]{8cm}
{The evolution of the fluctuation $\langle\chi^2\rangle$ 
in the case of $\phi_c=10^{-4}M_{\rm pl}$, $q=0.1$,
and for  $g=0.07$ (bottom), $g=0.1$
(middle), and $g=0.5$ (top) respectively.
The final variances are $\langle\chi^2\rangle_f=
1.2\times 10^{-10}M_{\rm pl}^2$, 
$2.5 \times 10^{-9}M_{\rm pl}^2$, and
$1.3 \times 10^{-6}M_{\rm pl}^2$ respectively.
}\\[1em]
\noindent
\parbox[t]{2cm}{FIG. 12:\\~}\ \
\parbox[t]{8cm}
{The evolution of the $\phi_0$ field in the case of 
$\phi_c=10^{-4}M_{\rm pl}$, $q=0.1$, and $g=0.5$.
While the back reaction effect of $\phi$ particles can
be negligible, the increase of $\langle\chi^2\rangle$
strongly affects the evolution of the $\phi_0$ field
from $\bar{t}\approx 0.17$.
}\\[1em]
\noindent
\parbox[t]{2cm}{FIG. 13:\\~}\ \
\parbox[t]{8cm}
{The evolution of $\langle\chi^2\rangle$ 
for $\phi_c=10^{-3}M_{\rm pl}$,
$q=0.1$, and $g=0.05$
in two cases of $m_{\chi}=3 \times 
10^{14}$ GeV (top) and 
$m_{\chi}=1 \times 10^{15}$ GeV (bottom).
The $\chi$ particle whose
mass is of order $m_{\chi}=10^{14}$ GeV is
generated, but the production of
the massive particle with 
$m_{\chi}~\mbox{\raisebox{-1.ex}{$\stackrel
     {\textstyle>}{\textstyle\sim}$}}~10^{15}$
GeV is hardly expected for $g=0.05$.
}\\[1em]
\parbox[t]{2cm}{FIG. 14:\\~}\ \
\parbox[t]{8cm}
{The evolution of $\langle\chi^2\rangle$ 
for $\phi_c=10^{-3}M_{\rm pl}$ and $q=0.1$, 
in two cases of $g=0.5$, $m_{\chi}=1 \times 
10^{15}$ GeV (top); and $g=1$,
$m_{\chi}=1 \times 10^{16}$ GeV (bottom).
In the case of $g=0.5$, the $\chi$ particle whose 
mass is of order $m_{\chi}=10^{15}$ GeV is
effectively enhanced.
In the case of $g=1$, the GUT scale boson 
$m_{\chi}=10^{16}$ GeV is a little produced
in the initial stage of the oscillating inflation.
}\\[1em]
\noindent


\begin{thebibliography}{99}
\bibitem{comment}
In this paper we do not consider the nonminimal coupling of
the $\phi$ or $\chi$ field with the scalar curvature $R$.
However, there is a possibility that the nonminimal coupling
would lead to the gravitational particle production 
as in Ref.~\cite{TMT}.
\bibitem{comment3}
This criterion was also obtained by analytic estimates performed
in Ref.~\cite{KLS2}.
\end{thebibliography}
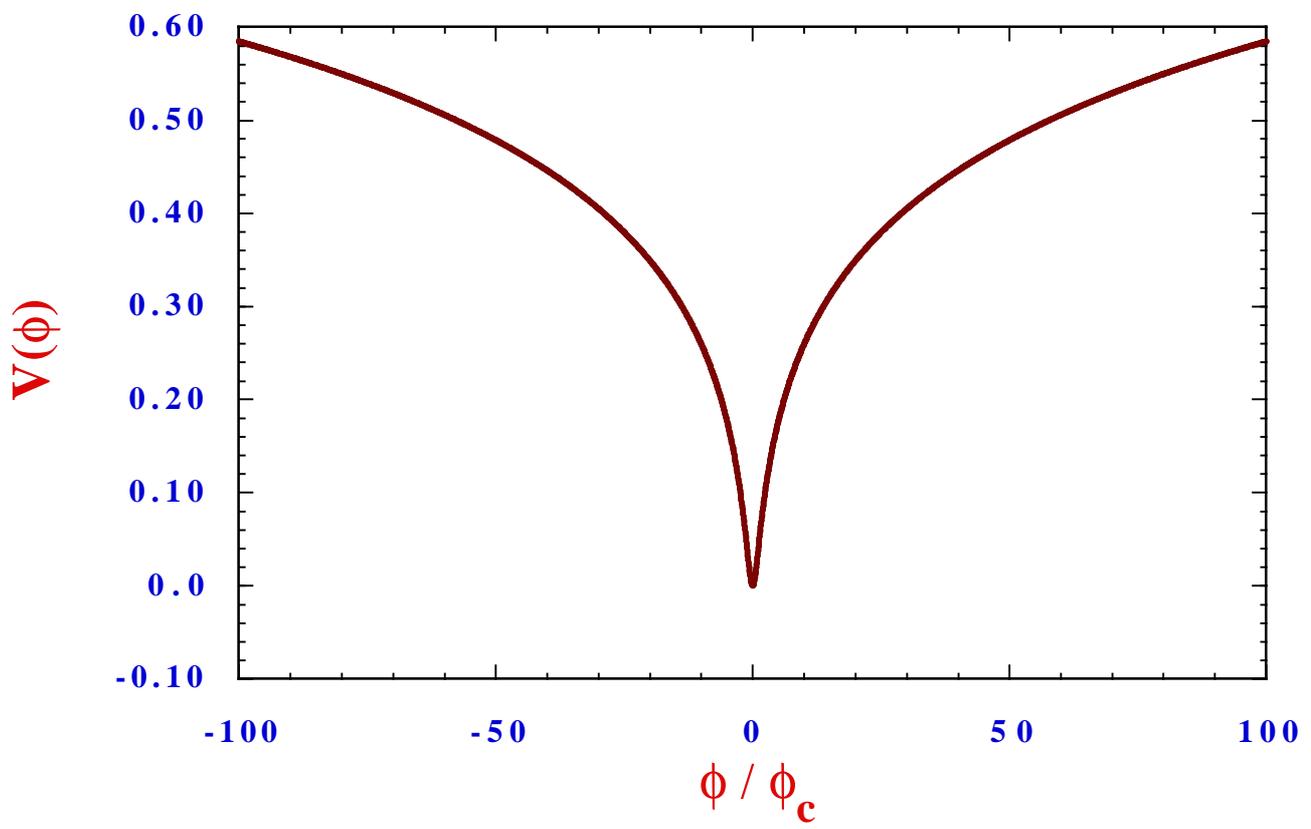
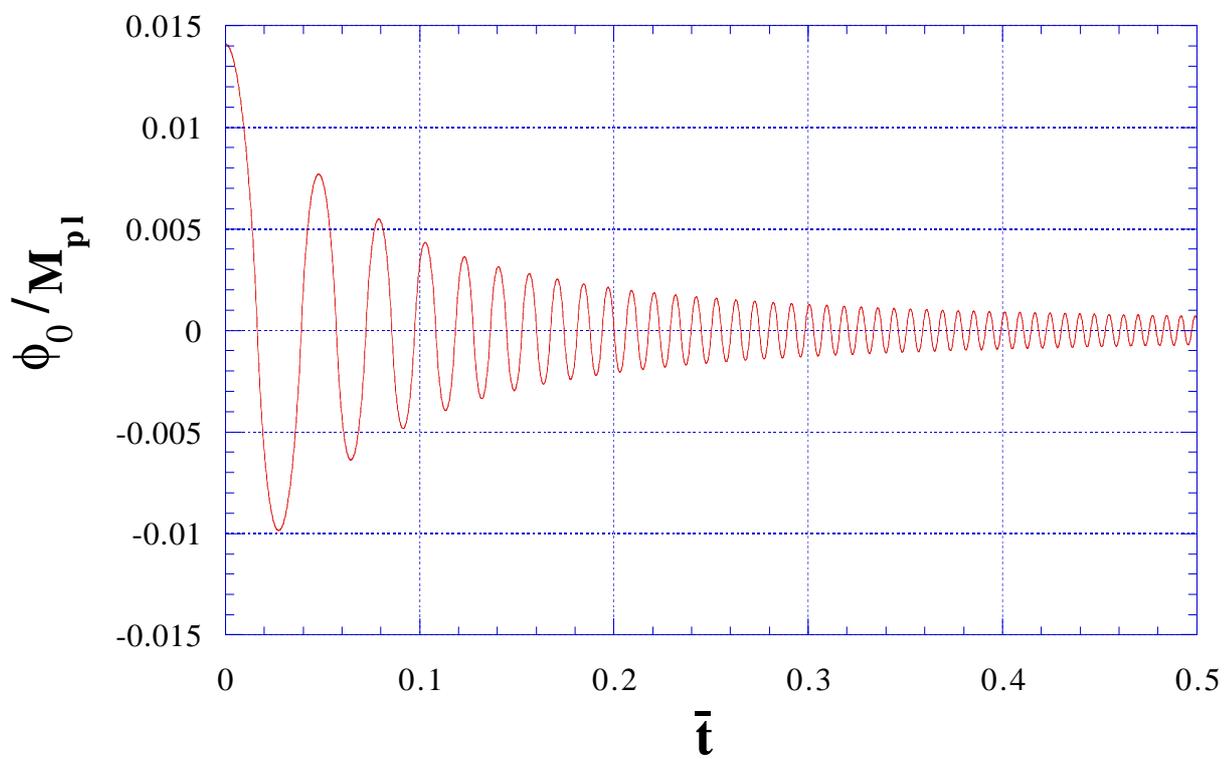
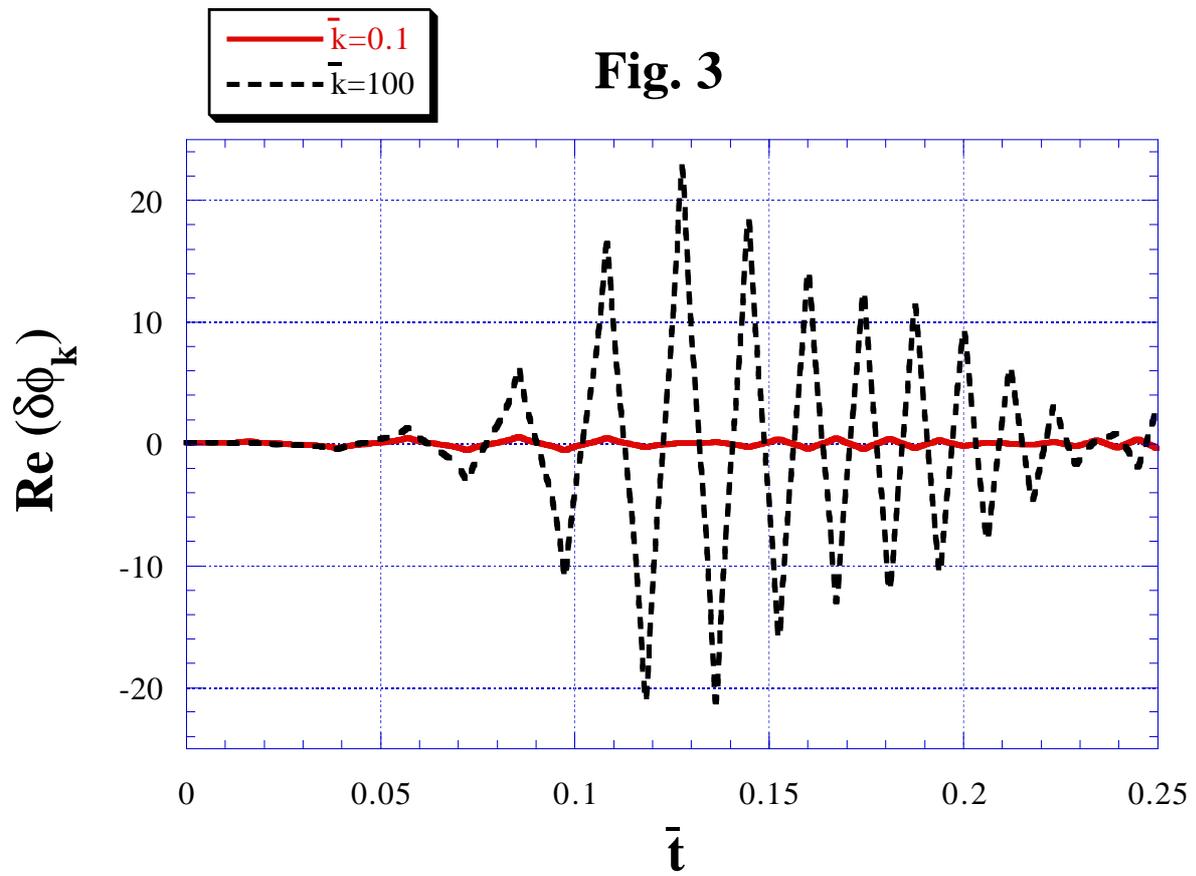
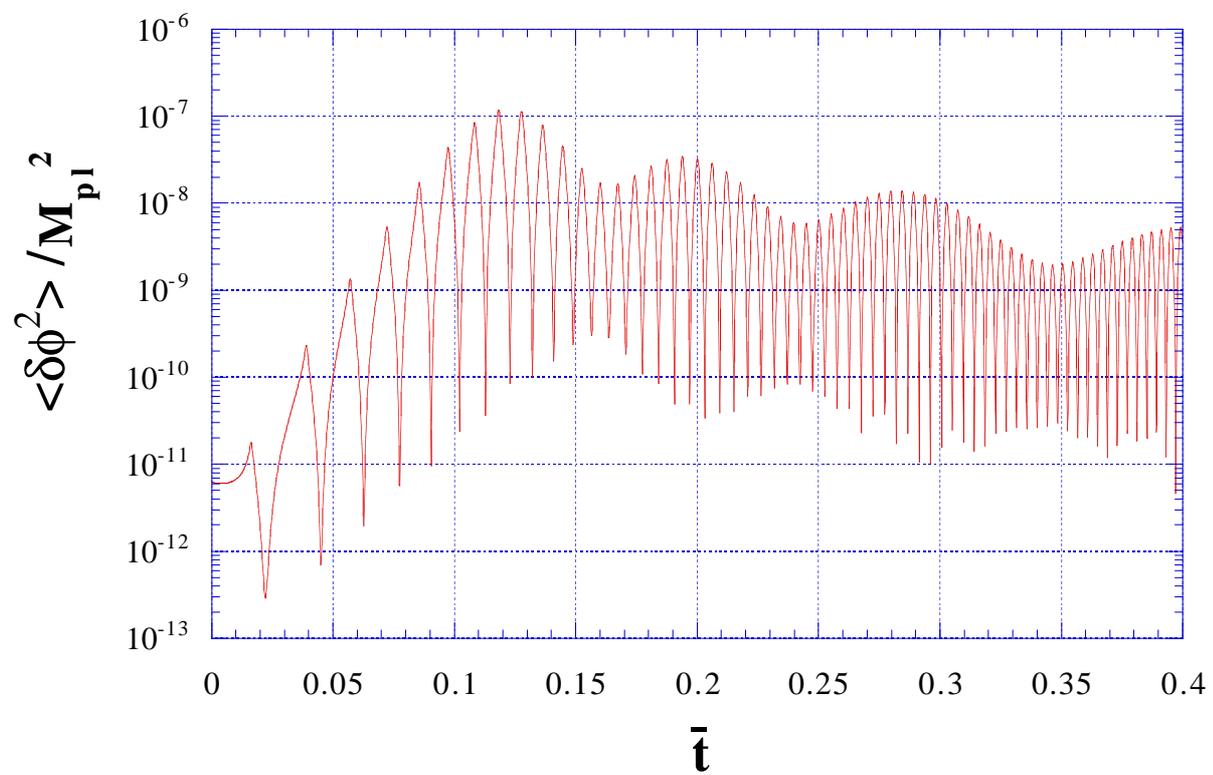
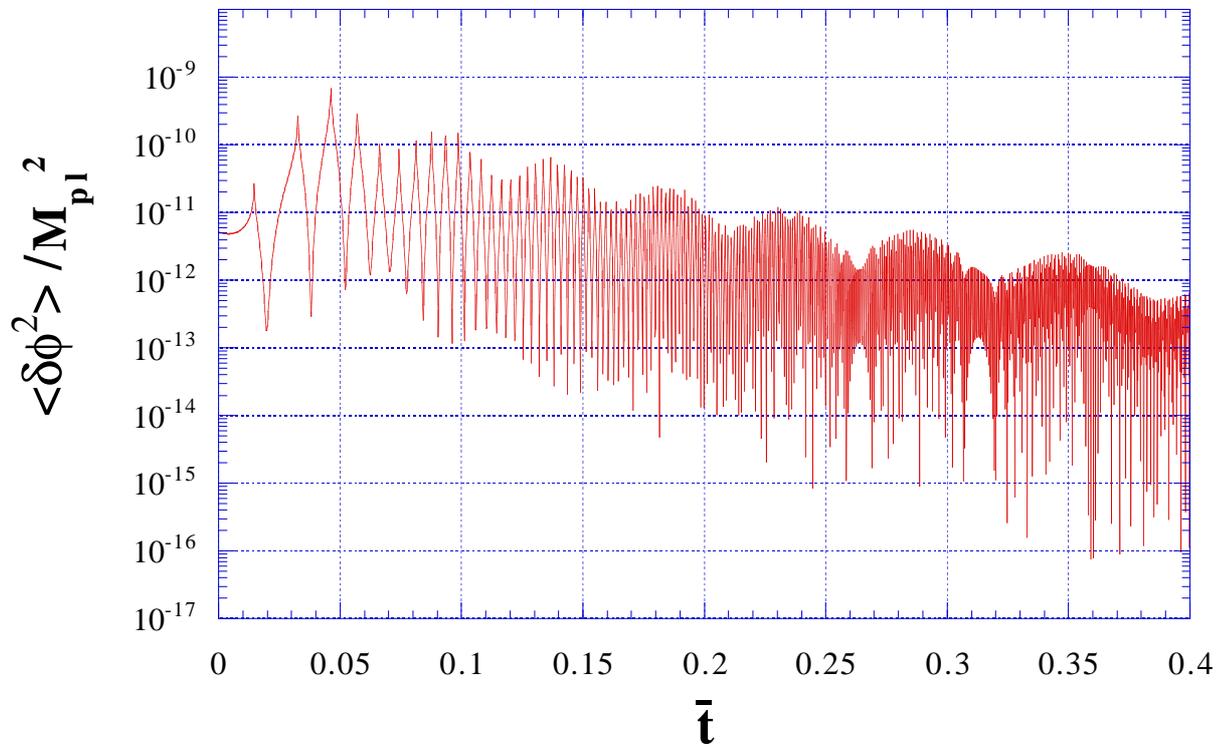
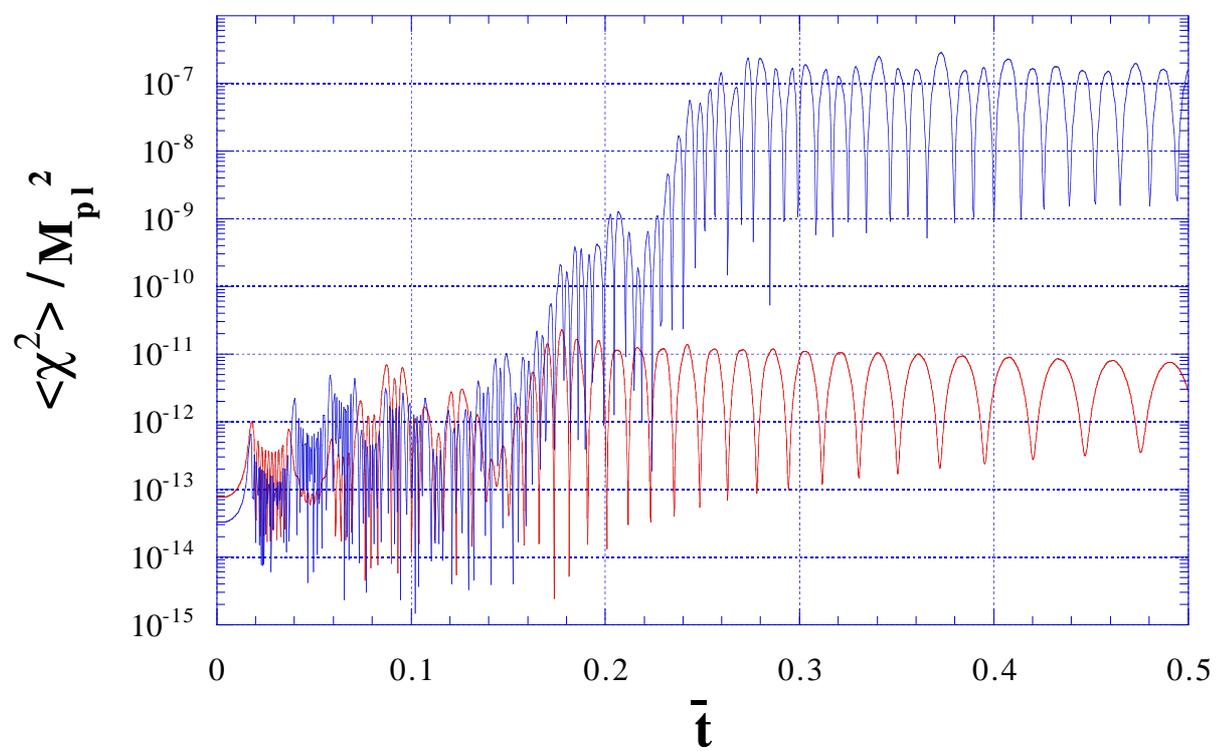
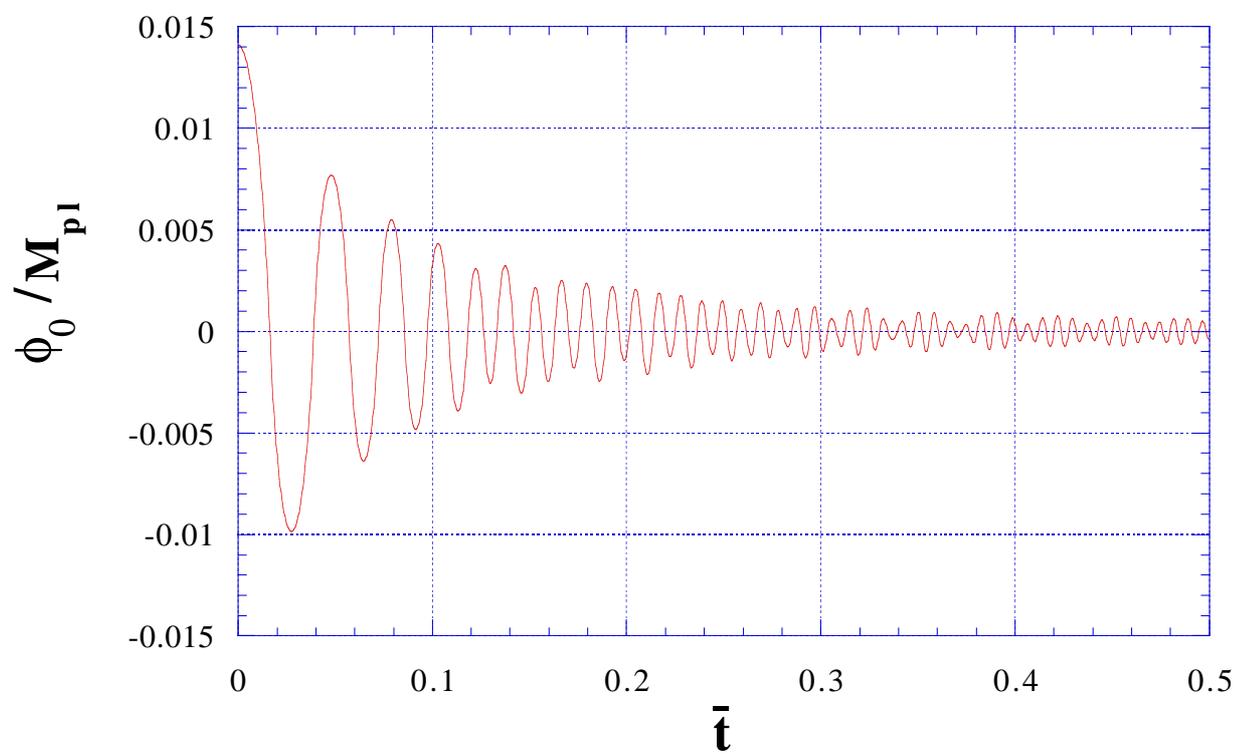
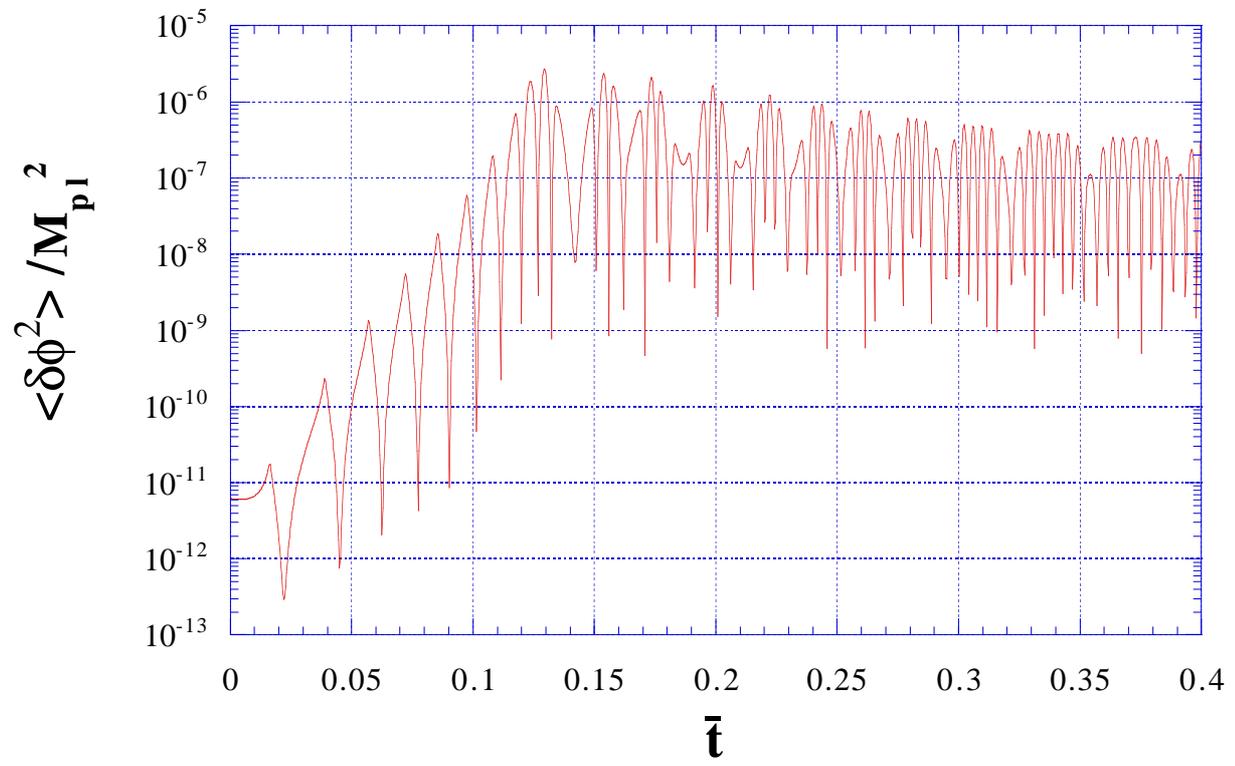
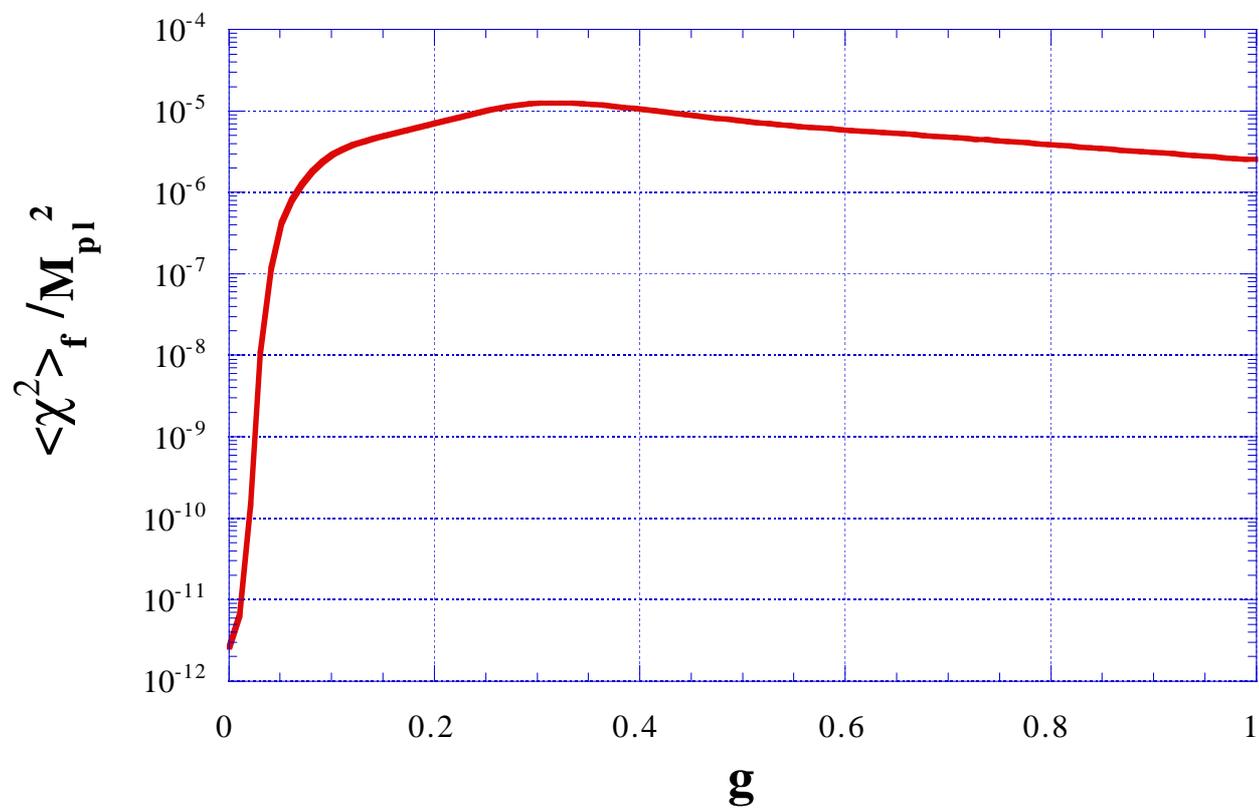
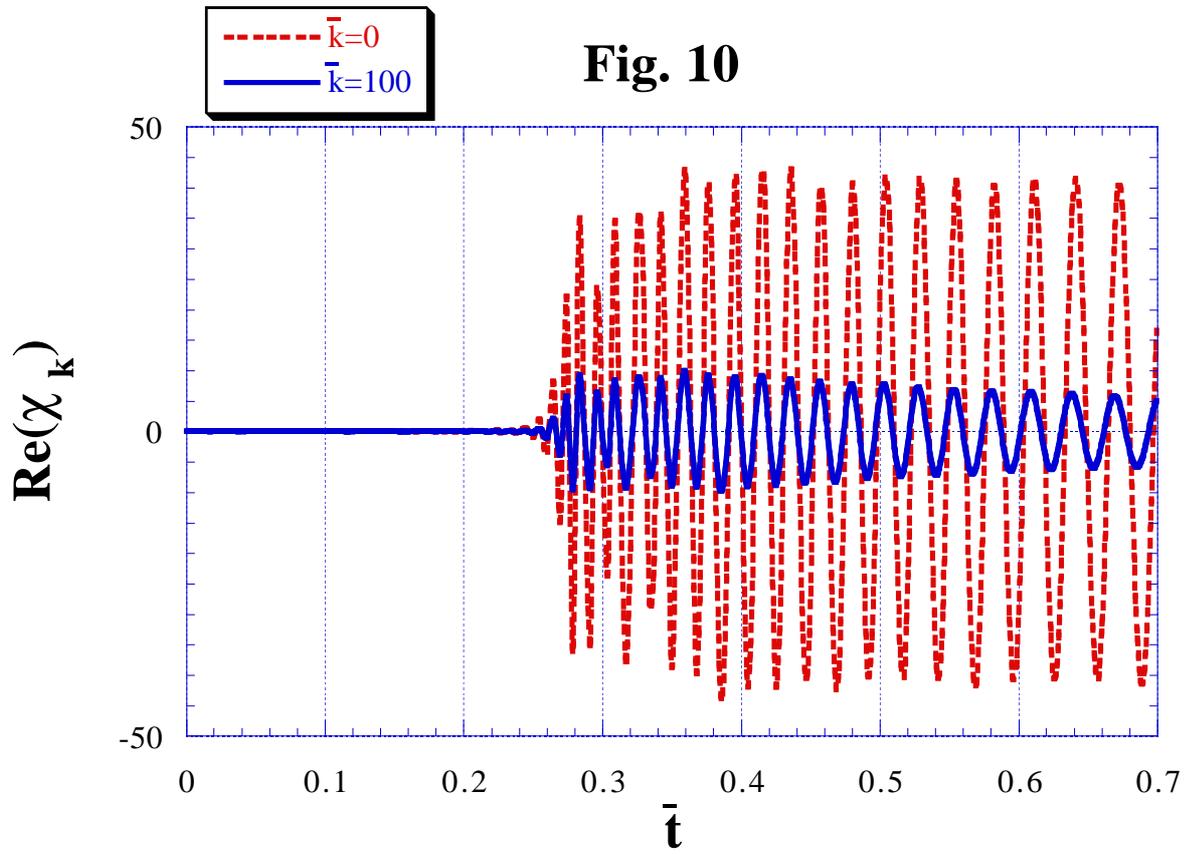
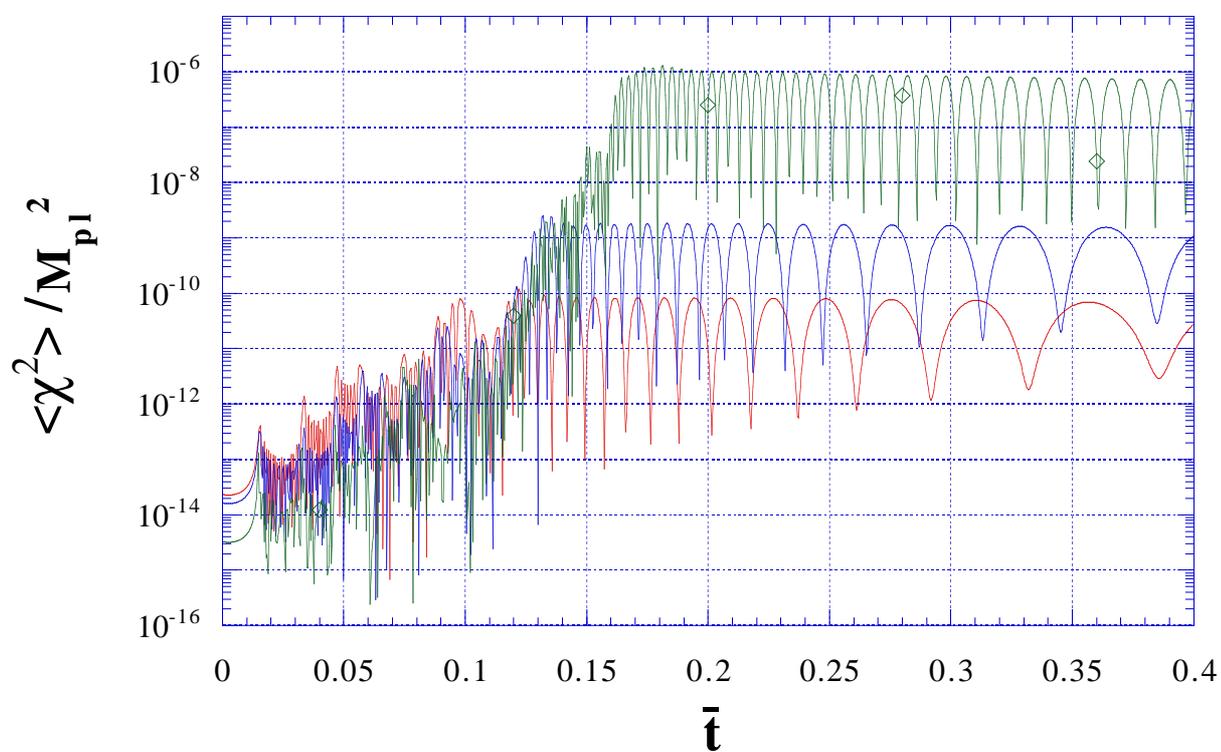
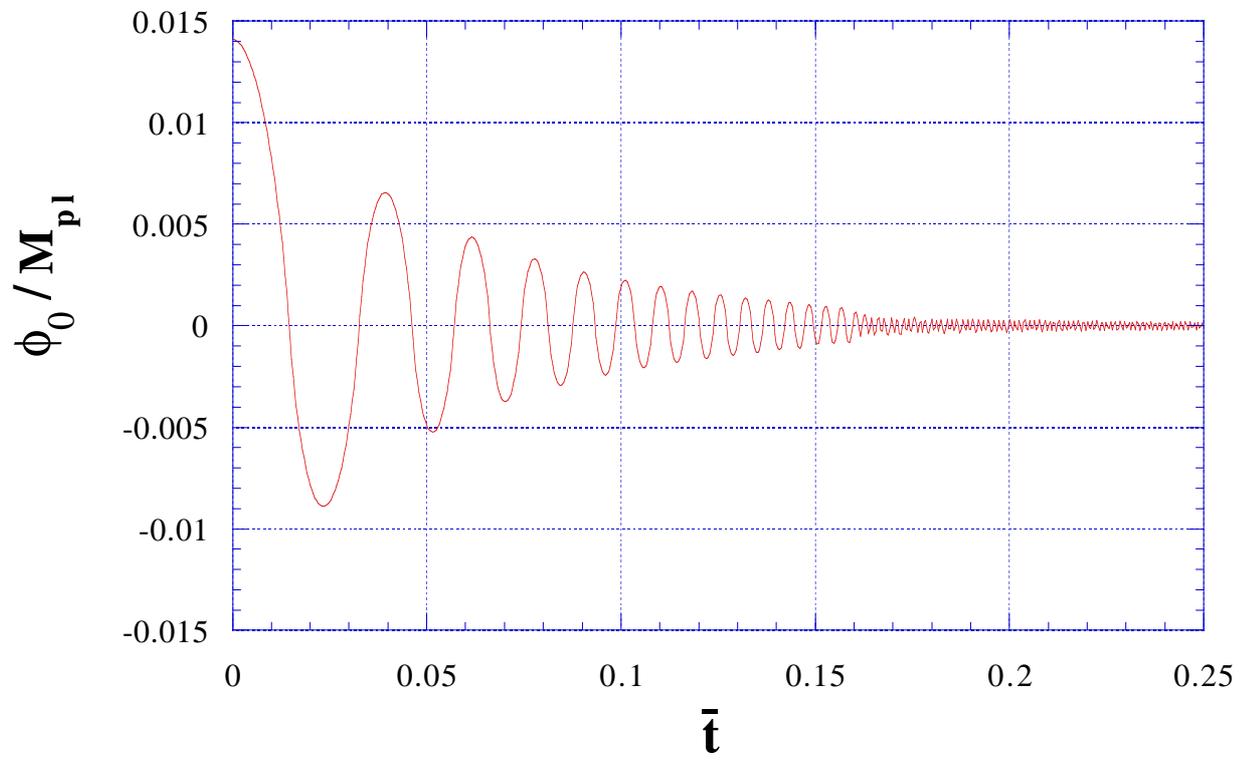
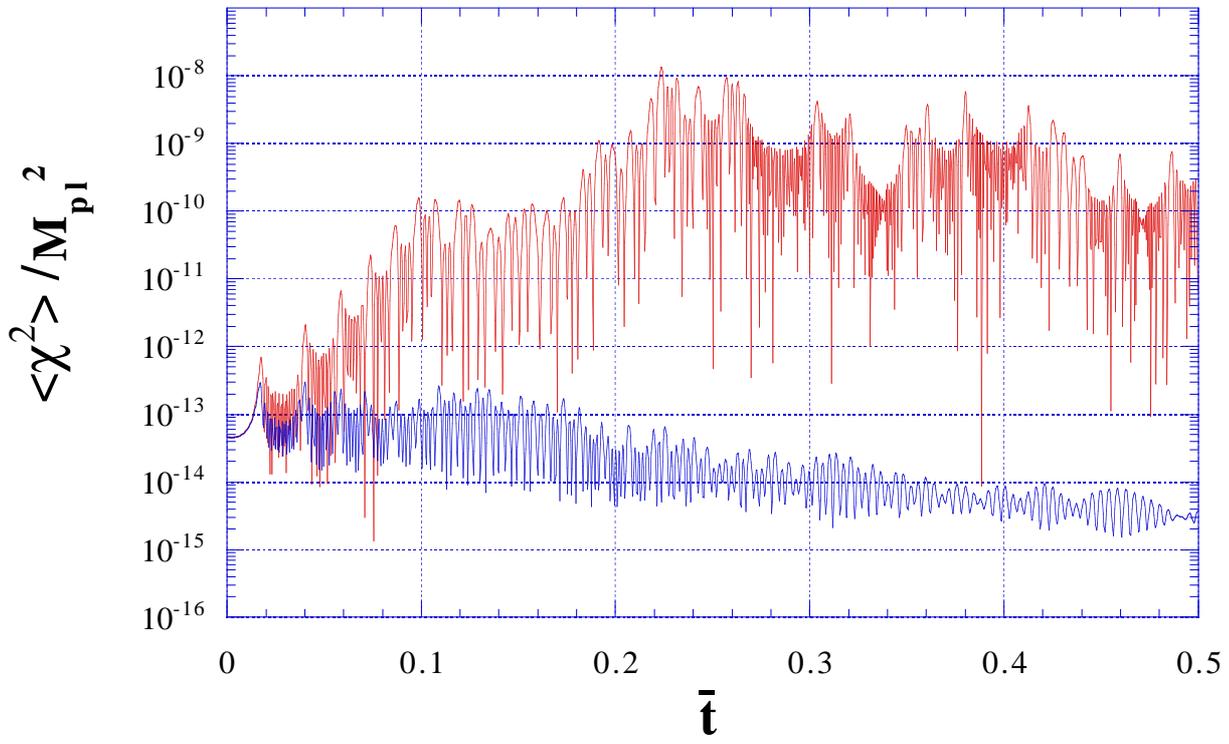
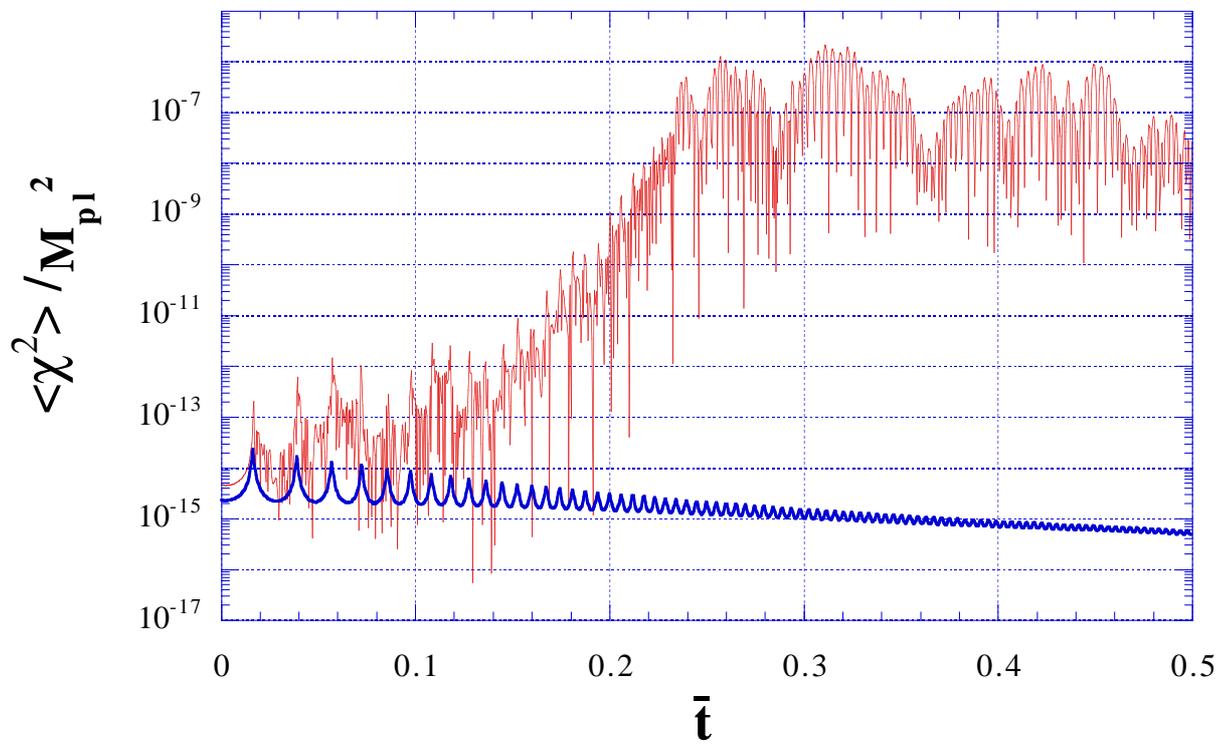
\end{document}